\documentclass[aps,prd,superscriptaddress,twocolumn,floatfix,showpacs]{revtex4}
\newcommand{\beq}{\begin{equation}}
\newcommand{\eeq}{\end{equation}}
\usepackage{graphicx}
\usepackage{epsf}
\begin{document}
\title{Fully General Relativistic Simulations of Black Hole-Neutron
  Star Mergers}

\author{Zachariah B. Etienne}
\email{zetienne@uiuc.edu}
\affiliation{Department of Physics, University of Illinois at
  Urbana-Champaign, Urbana, IL 61801}
\author{Joshua A. Faber}
\altaffiliation{National Science Foundation (NSF) Astronomy and
  Astrophysics Postdoctoral Fellow.}
\altaffiliation{Current Address: School of Mathematical Sciences,
  Rochester Institute of Technology, Rochester, NY 14623}
\affiliation{Department of Physics, University of Illinois at
  Urbana-Champaign, Urbana, IL 61801}
\author{Yuk Tung Liu}
\affiliation{Department of Physics, University of Illinois at
  Urbana-Champaign, Urbana, IL 61801}
\author{Stuart~L.~Shapiro}
\altaffiliation{Also at Department of Astronomy and NCSA, University of
  Illinois at Urbana-Champaign, Urbana, IL 61801}
\affiliation{Department of Physics, University of Illinois at
  Urbana-Champaign, Urbana, IL 61801}
\author{Keisuke Taniguchi}
\affiliation{Department of Physics, University of Illinois at
  Urbana-Champaign, Urbana, IL 61801}
\author{Thomas W. Baumgarte}
\altaffiliation{Also at Department of Physics, University of Illinois at
  Urbana-Champaign, Urbana, IL 61801}
\affiliation{Department of Physics and Astronomy, Bowdoin College,
  Brunswick, ME 04011}

\begin{abstract}
Black hole-neutron star (BHNS) binaries are expected to be among the
leading sources of gravitational waves observable by ground-based
detectors, and may be the progenitors of short-hard gamma ray bursts
(SGRBs) as well.  We discuss our new fully general relativistic
calculations of merging BHNS binaries, which use high-accuracy,
low-eccentricity, conformal thin-sandwich configurations as initial
data.  Our evolutions are performed using the moving puncture method
and include a fully relativistic, high-resolution shock-capturing
hydrodynamics treatment.  Focusing on systems in which
the neutron star is irrotational and the black hole is nonspinning
with a 3:1 mass
ratio, we investigate the inspiral, merger, and disk formation in the
system. We find that the vast majority of material is promptly
accreted and no more than 3\% of the neutron star's rest mass is
ejected into a tenuous, gravitationally bound disk.  We find similar
results for mass ratios of 2:1 and 1:1, even when we reduce the NS
compaction in the 2:1 mass ratio case.  These ambient disks reach
temperatures suitable for triggering SGRBs, but their masses may be
too small to produce the required total energy output.  We measure gravitational
waveforms and compute the effective strain in frequency space, finding
measurable differences between our waveforms and those produced by
binary black hole mergers within the advanced LIGO band. These
differences appear at frequencies corresponding to the emission that
occurs when the NS is tidally disrupted and accreted by the black
hole.  The resulting information about the radius of the neutron star
may be used to constrain the neutron star equation of state.
\end{abstract}


\maketitle

\section{Introduction}

Mergers of compact binaries, consisting either of neutron stars (NS)
or black holes (BH), are expected to be among the most promising
sources of gravitational waves detectable by ground-based laser
interferometers like LIGO \cite{LIGO1,LIGO2}, VIRGO
\cite{VIRGO1,VIRGO2}, GEO \cite{GEO}, and TAMA \cite{TAMA1,TAMA2}, as
well as by the proposed space-based interferometers LISA \cite{LISA} and
DECIGO \cite{DECIGO}.  Theoretical models indicate that a neutron
star-neutron star (NSNS) \cite{HMNS1,HMNS2,ShiUIUC,STU1,STU2,ST,SST}
or black hole-neutron star (BHNS) \cite{FBSTR,FBST,SU1,SU2,SST,ST07}
merger may result in a hot, massive disk around a BH, whose
temperatures and densities could be sufficient to trigger a short-hard
gamma-ray burst (SGRB).  Indeed, SGRBs have been repeatedly associated
with galaxies with extremely low star formation rates (see
\cite{SGRB_local} and references therein for a review), indicating
that the source is likely to involve an evolved population, rather
than main sequence stars.

Modeling the inspiral, coalescence and merger of compact binaries
requires fully general relativistic dynamical simulations, and has
been a long-standing goal of numerical relativity (see \cite{BS03} for
a review).  Historically, the first successful dynamical simulations
of compact binaries involved NSNS binaries~\cite{ShibFirstNSNS,ShiU00,
STU1,STU2,ST,mdsb04,mgs04,ahllmnpt07}).  
A breakthrough in the simulations of
BHBH binaries occurred more 
recently \cite{FP1,God1,RIT1}.  Simulations of BHNS binaries, on
the other hand, have so far lagged behind -- perhaps because they
combine the difficulties associated with black hole singularities
with the subtleties of relativistic hydrodynamics.  To date the only fully
self-consistent dynamical simulations of BHNS inspiral and
coalescence are those of Shibata and Ury\={u} \cite{SU1,SU2}
(hereafter SU) and Shibata and Taniguchi \cite{ST07} (hereafter ST).

Over the past years, we have systematically developed the tools
necessary to simulate the inspiral and merger of BHNS binaries,
including the tidal disruption of the neutron stars and the potential
formation of an accretion disk. 
As reviewed below, we have
constructed quasiequilibrium initial data describing relativistic BHNS binaries in
quasicircular orbits~\cite{BSS,TBFS05,TBFS06,TBFS07a,TBFS07b}, and
performed preliminary relativistic dynamical simulations by assuming
several simplifying approximations~\cite{FBSTR,FBST}.  We also
demonstrated and tested how the numerical techniques adopted in many recent BHBH
puncture simulations can be combined with relativistic hydrodynamics
to evolve quasiequilibrium initial data that are provided only in the
black hole exterior~\cite{FBEST,EFLSB}.  In this paper we report on
our first fully self-consistent, dynamical simulations of BHNS
binaries.

For the initial data, we have adopted a hierarchical approach to
construct quasiequilibrium models of BHNS in quasicircular orbit.  We
began by making a number of simplifying assumptions, and have relaxed
these assumptions step by step
(\cite{BSS,TBFS05,TBFS06,TBFS07a,TBFS07b}, cf.~\cite{Miller01,Grand06}
for other BHNS initial data).  Our current models, which we adopt as
initial data for the dynamical simulations described in this paper,
are solutions to Einstein's constraint equations in the conformal
thin-sandwich (CTS) decomposition.  We model the neutron star as an
irrotational $\Gamma=2$ polytrope, and impose the black hole
equilibrium boundary conditions of Cook and Pfeiffer \cite{CP04} on
the black hole horizon, to approximate an irrotational BH.  In our
most recent paper \cite{TBFS07b}, we adopted the methods of Caudill
{\it et.al.} \cite{Caudill} to construct irrotational
black holes more accurately, and found closer agreement with
post-Newtonian results.  These improved initial data will be
incorporated into our next set of dynamical calculations.

For the purpose of comparison we point out that the dynamical
simulations of SU and ST adopt initial data that are different from
ours. Both approaches lead to valid solutions to Einstein's
constraint equations, but the solutions may be physically distinct.
Specifically, our initial data use the CTS decomposition, 
which allows us to impose an approximate helical Killing vector 
on the spacetime and thereby set to
zero several time derivatives of the field variables in a corotating
frame. For example, we impose all conditions $\partial_t \tilde{\gamma}_{ij}=0$,
where $\tilde{\gamma}_{ij}$ is the conformally related spatial
metric, and these immediately yield a relation between the components of
the extrinsic curvature and the shift vector [see Eq.~(\ref{AijCTS})].  
By contrast, SU and ST do not impose these conditions, but instead 
use the conformal transverse-traceless (CTT) 
decomposition to obtain the extrinsic curvature on the initial
slice. However, they do employ the assumption of a helical Killing 
vector to construct a lapse and shift,  
(cf., \cite{TichyBrugLagun} who use a similar approach for BHBHs) 
and these gauge quantities are used to solve the quasiequilibrium
fluid equations for the neutron star.  They are also used to compute
the matter source terms appearing in the constraint equations.  For example, SU
and ST take the divergence of Eq.~(\ref{AijCTS}) to
generate three equations for the shift.  In addition, they model the
BH as a ``puncture'' (see \cite{BeiO94,BeiO96,BB97}), whereas we
excise the BH interior and impose the equilibrium boundary conditions
on the apparent horizon to force the BH to be stationary, at least
momentarily.  Therefore, one might speculate that our CTS
initial data may represent quasiequilibrium BHNSs in quasicircular
orbit more faithfully than SU and ST's initial data. 
In this paper, we present evidence that the details of
the initial data have a noticeable impact on the outcome of the merger
-- including the disk mass -- which may explain some of the
differences between the findings of SU and ST and ours.

In most dynamical simulations of BHNS binaries to date, the
self-gravity of the NS and/or the tidal gravity of the BH are
treated in a Newtonian or post-Newtonian framework (see, e.g.,
\cite{Lee00,RSW,Rosswog,Koba,RKLRasio}; see also \cite{LRA} who
performed fully relativistic simulations of head-on collisions).  In
many calculations, especially those with an initial mass ratio
$q=M_{\rm BH}/M_{\rm NS}\lesssim 3$, significant disks are formed
after the NS is disrupted, and for very stiff nuclear equations of
state (EOSs), the core of the NS may survive the initial mass transfer
episode and remain bound.  These findings contrast with some
semi-analytic relativistic arguments that suggest that it is very
difficult to form disks with appreciable masses in the merger of BHNS
binaries \cite{MCMiller}.

Using our earlier initial data \cite{BSS,TBFS05}, which assumed
extreme mass ratios with $q \gg 1$, we performed simulations of BHNS
merger in an approximate relativistic framework \cite{FBSTR,FBST}.  In
particular, we assumed that the spatial metric remains conformally
flat throughout the evolution (see \cite{Isen,WMM}).  Though this
approach only allows for crude estimates, we found that mergers of
irrotational BHNS binaries may lead to disks of masses up to 0.3
$M_{\odot}$, with sufficient heating to emit the neutrino fluxes that
are required to launch a gamma-ray burst. In their fully relativistic
BHNS simulations, SU later found disk masses in the range of 0.1 - 0.3
$M_{\odot}$ for corotating NSs, and ST found smaller disk masses of
0.04 - 0.16 $M_{\odot}$ for more realistic irrotational NSs.

In preparation for our fully relativistic dynamical simulations of
BHNS merger, we demonstrated in~\cite{FBEST,EFLSB} that the moving
puncture method (see \cite{God1,RIT1} as well as numerous later
publications), which has proven extremely useful for 
BHBH simulations, can be adopted for BHNS simulations and conformal
thin-sandwich initial data.  Two conceptional issues were
addressed, namely the inclusion of 
relativistic hydrodynamics into these simulations, and the fact that
moving puncture simulations require initial data everywhere, while our
conformal thin-sandwich initial data excise the black hole interior
and hence provide data only in the black hole exterior.

In contrast to the original dynamical puncture simulations
\cite{Bru99,BruTJ04}, in which the puncture was forced to remain at a
fixed coordinate location, the ``moving puncture'' approach allows the
punctures to move freely through the computational grid.  This method
is typically used in the context of the BSSN formulation \cite{SN,BS},
coupled to a ``Gamma-driving'' shift \cite{Gammadrive} and a
``1+log'' slicing condition \cite{BonMSS95}.  
Geometrical arguments show that, with this slicing condition, dynamical
simulations approach limit surfaces of finite areal radius around
black hole singularities, but never reach the singularity itself (see
\cite{JenaGeo1,JenaGeo2,Bro07,BN}).  These findings provide insight
into why moving puncture simulations can possibly be successful, and
also suggest that it may be possible to incorporate relativistic
hydrodynamics into these simulations.  Since the simulations only
cover regular regions of the spacetime, the hydrodynamic flow never
encounters any black hole singularities.  In \cite{FBEST} we
demonstrated that with only very minor modifications, our 
high-resolution shock-capturing (HRSC) relativistic hydrodynamics
algorithm (see \cite{DLSS}) can indeed be used together with the
moving puncture method to model accretion onto black holes.

Part of the appeal of the moving puncture approach stems from the fact
that it does not require an excision of the black hole interior.
Accordingly, this method requires initial data everywhere, both in the
BH's exterior and interior.  Most dynamical moving puncture
simulations to date have therefore adopted the puncture method also in
the construction of the initial data (see e.g., \cite{BB97,Bau00}).
As discussed above, CTS initial data are generally believed to be very
good approximations of true quasiequilibrium states, but solving the CTS
equations usually involves excising the BH interior, so that the
resulting data exist only in the BH exterior.  By definition, no
physical information can propagate from the BH interior to the
exterior, and we have recently demonstrated that even unphysical,
constraint-violating noise (``junk'') does not leave the BH
(\cite{FBEST,EFLSB}, see also \cite{Turducken}) in numerical
evolutions employing the BSSN formulation and moving puncture gauge
conditions (modulo certain smoothness restrictions on the junk data
near the horizon).  Thus, the BH interiors in CTS initial data can be
filled with ``junk'' without affecting the external spacetime,
enabling us to evolve our quasiequilibrium BHNS initial data via the
moving puncture formalism.

With all the aforementioned pieces in place, we now report our first
fully self-consistent, relativistic dynamical simulations of BHNS
binaries.  We are particularly interested in binaries that may
potentially lead to a sizable accretion disk, so we focus on binaries
in which the neutron star is tidally disrupted just before reaching the
innermost stable circular orbit (ISCO) and plunging into the black
hole.  The binary separation $d_{\rm tid}$ at which the neutron star
will be tidally disrupted may be estimated from the following crude
Newtonian argument.  Equating the tidal force exerted by the BH on a
test mass at the NS's surface with the gravitational force exerted by
the neutron star on this test mass, we find
\begin{equation} \label{eq:d_tid}
\frac{d_{\rm tid}}{M_{\rm BH}} \simeq q^{-2/3} {\cal C}^{-1},
\end{equation}
where ${\cal C}\equiv M_{\rm NS}/R_{\rm NS}$ is the neutron star
compaction.  Given typical neutron star compactions of ${\cal C} \sim
0.2$, small but reasonable values of $q$ are required for $d_{\rm
tid}$ to be larger than the ISCO separation of about $d_{\rm ISCO}
\sim 6 M_{\rm BH}$.  Our more careful analysis (\cite{TBFS07b}; see
Fig.~15) shows that for $\Gamma = 2$ polytropes we need $q \lesssim
4.25$ .  Given typical neutron star masses ($M_{\rm NS} \sim 1.5
M_{\odot}$), this means that we can expect the formation of an
accretion disk only for low-mass black holes.
 
How often such binaries merge in the observable universe is still an
open question.  The uncertainties arise from some aspects of
population synthesis calculations that are only partially understood.
In particular, envelope ejection efficiency during the common envelope
phase seems to be a crucial factor in forming low-mass BHs during
binary stellar evolution.  For example, if one assumes efficient
ejection and a large maximum NS mass, the primary NS will generally
accrete insufficient mass to induce collapse to a BH, and one ends up
with a large number of NSNS binaries.  For inefficient ejection and a
smaller maximum NS mass, it is relatively easy for the NS to accrete
sufficient material to form a BH with a mass only slightly larger than
a NS.  Although some previous population synthesis calculations
working under the latter assumption found a nearly flat spectrum of
binary mass ratios spanning the range $q=1.5-10$ \cite{BKB}, a more
recent calculation that assumes highly efficient envelope ejection
yields typical binary mass ratios $q=6-10$ \cite{BTRvdS}.  The latter
scenario would predict that NSs undergoing tidal breakup prior to
reaching the ISCO are rare, as are any resulting SGRBs from these
systems.  The overall rate estimates for BHNS mergers observable by an
advanced LIGO detector typically fall in the range ${\cal R}\sim
1-100~{\rm yr}^{-1}$ \cite{KBKOW}.

These issues noted, we begin our investigation of BHNS binary merger
and coalescence in full general relativity.  This paper is the first
in a sequence of papers which will thoroughly explore the effect of
various binary parameters on the tidal disruption, disk formation, the
potential for launching a GRB, and the corresponding gravitational
wave signals.  In this paper we will focus on irrotational binaries
with mass ratios $q \leq 3$.  As an additional word of caution, we
point out that our results are fundamentally limited by uncertainties
about the true nuclear EOS, both in the cold initial state as well
as the later shock-heated hot phase.  Disk masses may depend
sensitively on the structure of the NS, especially the low-density
outer regions, so all BHNS merger results should be viewed in light of
this caveat.  In particular, the likelihood of BHNS mergers as SGRB
progenitors may be difficult to determine conclusively until this
issue is resolved.

This paper is organized as follows.  In Secs.~\ref{sec:basic_eqns} and
\ref{sec:numerical}, we summarize the basic equations and their
specific implementation in our general relativistic hydrodynamics
scheme, along with a discussion of initial data, gauge conditions,
matter evolution, and diagnostics.  In Sec.~\ref{sec:runs}, we discuss
the results of our BHNS merger simulations, and how they depend on
both physical as well as computational parameter choices.  We conclude
in Sec.~\ref{sec:discussion} with a discussion of our findings, and
comment on future directions.

\section{Basic Equations}\label{sec:basic_eqns}

In this section we list the full set of evolution equations integrated
by our numerical code.  Field, coordinate, and hydrodynamic evolution
equations are summarized in Secs.~\ref{sec:bssn},~\ref{sec:gauge},
and~\ref{sec:hydro}, respectively.

\subsection{Field Evolution: The BSSN Equations}\label{sec:bssn}
Assuming geometrized units in which $G = c = 1$, we write the
spacetime metric in the standard 3+1 form
\begin{equation} %DMSB (1)
ds^2 = -\alpha^2 dt^2 + \gamma_{ij}(dx^i+\beta^idt)(dx^j+\beta^jdt),
\end{equation}
where $\alpha$, $\beta^i$, and $\gamma_{ij}$ are the lapse, shift, and
spatial 3-metric, respectively. The extrinsic curvature $K_{ij}$ is defined by
\begin{equation}%DMSB (2)
\label{Kij}
(\partial_t - {\mathcal{L}}_{\beta})\gamma_{ij} = -2\alpha K_{ij}.
\end{equation}
Here ${\mathcal{L}}_{\beta}$ is the Lie derivative with respect to
$\beta^i$.

In the BSSN formalism,
we define the conformally related metric $\tilde\gamma_{ij}$, the
conformal exponent $\phi$, the trace of the extrinsic curvature $K$,
the conformal traceless extrinsic curvature $\tilde A_{ij}$, and the
conformal connection functions $\tilde\Gamma^i$ as follows
\begin{eqnarray}
\label{YLsplit}  %DMSB (8)-(10)
  \phi &=& \frac{1}{12} \ln [ \det(\gamma_{ij})], \\
  \tilde\gamma_{ij} &=& e^{-4\phi} \gamma_{ij}, \\
  K &=& \gamma_{ij} K^{ij}, \\
   \tilde A_{ij} &=& e^{-4\phi} \left( K_{ij} - \frac{1}{3} \gamma_{ij} K
\right), \\
\tilde\Gamma^i &=& -\tilde\gamma^{ij}{}_{,j},
\end{eqnarray}
where $_{,j}$ denotes the partial derivative:
$\tilde\gamma^{ij}{}_{,j}\equiv\partial_j\tilde\gamma^{ij}$.
We use the same field evolution equations as Eqs.~(11)--(15) of
\cite{DMSB}:
\begin{eqnarray}
\label{evolve_gamma}  %DMSB (11)-(14)
(\partial_t - {\mathcal{L}}_{\beta})\tilde\gamma_{ij}
                 &=& -2\alpha\tilde A_{ij}, \\
\label{evolve_phi}
(\partial_t - {\mathcal{L}}_{\beta})\phi
                 &=& -{1\over 6}\alpha K, \\
\label{evolve_K}
(\partial_t - {\mathcal{L}}_{\beta})K
                 &=& -\gamma^{ij}D_jD_i\alpha + {1\over 3}\alpha K^2 \\
                 & & + \alpha \tilde A_{ij}\tilde A^{ij}
                     + 4\pi\alpha (\rho + S), \nonumber \\
\label{evolve_A}
(\partial_t - {\mathcal{L}}_{\beta})\tilde A_{ij}
                 &=& e^{-4\phi}(-D_iD_j\alpha
                     + \alpha(R_{ij}-8\pi S_{ij}))^{TF} \nonumber \\
                 & & + \alpha(K\tilde A_{ij} - 2\tilde A_{il}\tilde 
A^l{}_j),
\end{eqnarray}
and
\begin{eqnarray}
\label{evolve_Gamma} %DMSB (15)
\partial_t\tilde\Gamma^i &=& \partial_j(2\alpha\tilde A^{ij}
                 + {\mathcal{L}}_{\beta}\tilde\gamma^{ij}) \nonumber \\
         &=& \tilde\gamma^{jk}\beta^i{}_{,jk}
                 + {1\over 3}\tilde\gamma^{ij}\beta^k{}_{,kj}
  - \tilde\Gamma^j\beta^i{}_{,j} \\
  & &+{2\over 3}\tilde\Gamma^i\beta^j{}_{,j}
     + \beta^j\tilde\Gamma^i{}_{,j} - 2\tilde A^{ij}\partial_j\alpha
         \nonumber \\
  & &- 2\alpha\left({2\over 3}\tilde\gamma^{ij}K_{,j} - 6\tilde 
A^{ij}\phi_{,j}
     - \tilde\Gamma^i{}_{jk}\tilde A^{jk} + 8\pi\tilde\gamma^{ij}S_j
         \right).
         \nonumber
\end{eqnarray}

\subsection{Gauge Equations}\label{sec:gauge}

As in most moving puncture calculations, we use an advective
``1+log'' slicing condition for the lapse
\begin{equation}  %See any puncture paper ever
\partial_t\alpha-\beta^i\partial_i\alpha=2\alpha K,\label{eq:lapseevol}
\end{equation}
and a second-order ``non-shifting-shift'' (in the language of
\cite{GodGauge}) 
\begin{eqnarray}  %See \cite{FBEST}
\partial_t\beta^i &=& \frac{3}{4} B^i,\label{eq:shiftevol1}\\
\partial_t B^i&=&\left(\frac{dr}{d\bar{r}}\right)^2 \partial_t
{\tilde{\Gamma}}^i -\eta B^i.\label{eq:shiftevolfish}
\end{eqnarray}
This condition \cite{FBEST} is similar to that in
\cite{RIT1}, but allows for a fisheye radius $\bar{r}$, discussed
in Sec.~\ref{sec:num_grid_id} below.  This expression is closely
related to the ``Gamma-driver'' family of shift evolution equations.
We have found empirically that setting $\eta\sim0.5/M$ yields
well-behaved coordinate evolutions, where $M$ is the ADM mass
of the system, as defined in Eq.~(\ref{eq:madm}).  This value is chosen
for all runs except run B, for which we use $\eta=0.413/M$.

%This gauge condition differs in practice from that used in
%\cite{SU1,SU2}, in that they add an explicit damping term to their
%Gamma-driver expression of the form $\Delta t \partial_t F$, which
%essentially evaluates the expression $\tilde{\Gamma}^i$ one timestep
%forward in time.

\subsection{Hydrodynamic Equations}\label{sec:hydro}

The matter source terms are defined as
\begin{eqnarray} %DLSS (8)
   \rho &=& n_{\alpha}n_{\beta}T^{\alpha\beta}\ , \nonumber \\
   S_i  &=& -\gamma_{i\alpha}n_{\beta}T^{\alpha\beta}\ , \label{source_def} 
\\
   S_{ij} &=& \gamma_{i\alpha}\gamma_{j\beta}T^{\alpha\beta}\ ,\nonumber
\end{eqnarray}
where $T^{\alpha\beta}\equiv (\rho_0 + \rho_0\epsilon +
P)u^{\alpha}u^{\beta} + P g^{\alpha\beta}$ is the stress-energy tensor
for a perfect fluid,
$\rho_0$, $\epsilon$, $P$ and $u^\alpha$ are the fluid's rest-mass
density, specific internal energy, pressure, and 4-velocity, respectively,
and $n_{\alpha} = (-\alpha,0,0,0)$ is the future-directed unit normal to
the time slice.

We evolve the ``conserved hydrodynamic variables'', $\rho_*$,
$\tilde{S}_i$ and $\tilde{\tau}$, defined as follows
\begin{eqnarray} %See DLSS
\label{eq:rhosdef}\rho_* &=& -n_{\mu} \rho_0 u^{\mu}
= \alpha \sqrt{\gamma} \rho_0 u^0,\\
\label{eq:momdef}
   \tilde{S}_i &=& \sqrt{\gamma} S_i = \alpha \sqrt{\gamma} T^0{}_i=\rho_* 
h  u_i, \\
\label{eq:taudef}\tilde{\tau} &=& \sqrt{\gamma}\, n_{\mu} n_{\nu} T^{\mu 
\nu} - \rho_*
  = \alpha^2 \sqrt{\gamma}\, T^{00} - \rho_* .
\end{eqnarray}
The evolution equations for these variables are given by
Eqs.~(34), (36), and (38) of \cite{DLSS},
\begin{eqnarray} %DLSS (34), (36), (38)
\partial_t \rho_* + \partial_j (\rho_* v^j) &=& 0,\\
\partial_t \tilde{S}_i
  + \partial_j (\alpha \sqrt{\gamma}\, T^j{}_i) &=& \frac{1}{2} \alpha 
\sqrt{\gamma}
\, T^{\alpha \beta} \partial_i g_{\alpha \beta},\\
\partial_t \tilde{\tau} + \partial_i ( \alpha^2 \sqrt{\gamma}\, T^{0i}
-\rho_* v^i) &=& s,
\end{eqnarray}
where $\gamma \equiv \det(\gamma_{ij}) = e^{12\phi}$, and the
energy source term $s$ is
\begin{eqnarray} %DLSS (39)
   s &=& -\alpha \sqrt{\gamma}\, T^{\mu \nu} \nabla_{\nu} n_{\mu}  \cr
    &=& \alpha \sqrt{\gamma}\, [ (T^{00}\beta^i \beta^j + 2 T^{0i} \beta^j
+ T^{ij}) K_{ij} \cr
  & & - (T^{00} \beta^i + T^{0i}) \partial_i \alpha ].
\end{eqnarray}
Here $v^i\equiv u^i/u^0$ is the fluid's 3-velocity.

To complete the system of equations, we specify an EOS. Our code is
capable of handling EOSs of the form $P=P(\rho_0,\epsilon)$. In
this paper, we employ the standard $\Gamma$-law EOS
\begin{equation} \label{eq:eos}
P=(\Gamma-1)\rho_0 \epsilon
\end{equation}
with $\Gamma=2$ to model the NS matter.

\section{Numerical Methods}\label{sec:numerical}

The code we use is very similar to that described in
\cite{FBEST,EFLSB}.  We do not consider magnetic fields in this paper,
so the magnetic field sector is disabled in these calculations.  The
equations of general relativistic (GR) hydrodynamics are handled by a
HRSC technique~\cite{DLSS} that employs the monotonized central (MC)
reconstruction scheme~\cite{vL77} coupled to the HLL (Harten, Lax, and
van Leer) approximate Riemann solver~\cite{HLL}.  The metric is
evolved via the BSSN formalism~\cite{BS,SN} as described in
\cite{DMSB}, but with fourth-order accurate spatial differencing and
upwinding on the shift advection terms.  Our code is based on the
Cactus parallelization framework \cite{Cactus}, in which our
second-order Iterated Crank-Nicholson time-stepping is managed by the
{\tt MoL}, or method of lines, thorn.

For completeness, we provide below a brief overview of our grid setup and
initial data (Sec.~\ref{sec:num_grid_id}); a discussion of field,
gauge (Sec.~\ref{sec:num_gauge_field}), and hydrodynamic
(Sec.~\ref{sec:num_hydro}) evolution techniques; a description of how
we apply boundary conditions (Sec.~\ref{sec:num_bc}); a summary of
diagnostic techniques we use to both validate our numerical results and
examine our spacetimes (Sec.~\ref{sec:num_diags}); and finally a
description of the technique we use to measure gravitational wave (GW)
emission (Sec.~\ref{sec:num_gw}).

\subsection{Grid Setup and Initial data}\label{sec:num_grid_id}

We use a ``fisheye'' coordinate system \cite{RIT2} to expand the
physical extent of our numerical grid while maintaining high
resolution in the strong-field region.  Significantly lower resolution
is maintained in the wavezone, but it is set so that a gravitational
wavelength is resolved by at least 12 gridpoints.  To set up a fisheye
coordinate grid, we define the ``physical'' radius $r$ in terms of a
fisheye radius $\bar{r}$ according to
\begin{equation}
   r=a_n\bar{r}+\sum_{i=1}^{n}\frac{(a_{i-1}-a_i)s_i}{2\tanh(R_i/s_i)}
\ln\left(\frac{\cosh((\bar{r}+R_i)/s_i)}{\cosh((\bar{r}-R_i)/s_i)}\right).\label{eq:fishcoord}
\end{equation}
Here $a_i$ sets the magnitude of the $i$'th fisheye transition, $s_i$
determines the width of the transition, and $R_i$ specifies the center
of the transition.  In physical coordinates, the grid spacing smoothly
transforms from $\Delta x\approx a_{i-1}\Delta\bar{x}$ to $\Delta
x\approx a_i\Delta\bar{x}$, over a set of fisheye coordinates spanning
radii $R_i-s_i <\bar{r}<R_i+s_i$.  For convenience, we always set
$a_0=1$, so that our coordinate grid spacing $\Delta x$ represents the
physical grid spacing in the central region of our numerical grid. 
In this paper we use only one transition zone ($n=1$), with $a_1=8$.
We have listed other relevant grid parameters in
Table~\ref{table:grid}.

We defer to Appendix A of \cite{FBEST} for transformation laws
pertaining to all field and hydrodynamic quantities under a fisheye
transformation.  All of our calculations are performed assuming
equatorial symmetry, on numerical grids of the form $2N\times 2N\times
N$, with $N$ ranging from 166 to 305.

All initial data we evolve in this paper were generated by
\cite{TBFS07a}.  To map these spectral configurations onto our
non-spectral simulation grid, we first construct our numerical grid
and record the positions of each point in physical coordinates.  Then
we evaluate the field and hydrodynamic quantities based on their
spectral coefficients.  Next, we transform the vector and tensor
quantities into fisheye coordinates via transformations found in
Appendix~A of \cite{FBEST}.  Finally, the excised BH region is filled
with constraint-violating initial data, using the ``smooth junk''
technique we developed and validated in \cite{EFLSB}.

The assumptions
under which our initial data are constructed differ from those of SU
and ST.  We solve Einstein's constraint equations in the conformal
thin-sandwich (CTS) formalism, which allows us to impose an
approximate helical Killing vector by setting the time
derivatives of the conformally related metric to zero.  As a result, our
extrinsic curvature is always related to the shift vector that appears
in the solution through Eq.~(4) of \cite{TBFS07b},
\begin{equation}
\tilde{A}^{ij}\equiv \frac{\psi^6}{2\alpha}\left(\nabla^i\beta^j +
\nabla^j \beta^i
-\frac{2}{3}\tilde{\gamma}^{ij}\nabla_k\beta^k\right).\label{AijCTS}
\end{equation}
These quasiequilibrium initial data excise the black hole interior, 
allowing us to impose equilibrium boundary conditions on the excision surface. 
By contrast, SU and ST adopt the CTT decomposition to obtain
the extrinsic curvature on the initial slice, but employ the assumption of 
a helical Killing vector to construct a lapse and shift.
Also, they model the black hole as a puncture (see
\cite{BeiO94,BeiO96,BB97,TichyBrugLagun}). 
Both sets of initial data are solutions
to Einstein's constraint equations, but they may differ in both the
amount of spurious gravitational wave content and the degree of
orbital eccentricity.

\subsection{Metric Evolution and Gauge}\label{sec:num_gauge_field}

We apply two methods that have improved the stability and accuracy of our
field and gauge evolution when evolving BH spacetimes.

First, we use fourth-order finite differencing schemes to calculate
spatial derivatives in the field/gauge evolution sectors.  Also, for
any terms of the form $\beta^i\partial_i \ldots$, which arise in both
the Lie derivative terms and in the lapse evolution, we use fourth
order upwind differencing stencils instead of the standard centered
fourth-order stencils (see Eqs.~(2.5)-(2.6) and (2.2)-(2.4) of
\cite{ZBCL}, respectively).  We note for completeness that our mixed
second-derivative stencil is slightly different that given by Eq.(2.4)
of \cite{ZBCL}, but remains fourth-order convergent,
\begin{eqnarray}  % See Eq.(2.4) of \cite{ZBCL} -- I can demonstrate
                  % the math for this one
\partial_{xy}F_{i,j,k}=&&\frac{1}{48dx~dy}\left[F_{i-2,j+2,k}+F_{i+2,j-2,k}\right.\\
&& -F_{i+2,j+2,k}-F_{i-2,j-2,k} \nonumber\\
&& +16(F_{i+1,j+1,k}+F_{i-1,j-1,k} \nonumber\\
&&\left. -F_{i-1,j+1,k}-F_{i+1,j-1,k}) \nonumber\right].
\end{eqnarray}

In addition, we enforce the conditions $\tilde{\gamma} \equiv
\det(\tilde{\gamma_{ij}})=1$ and $\tilde{A} \equiv {\rm
tr}~(\tilde{A}_{ij})=0$ at every timestep, using the substitutions
\begin{eqnarray}  %These are rather obvious
\tilde{\gamma}_{ij}&\rightarrow&\tilde{\gamma}_{ij}/\tilde{\gamma}^{1/3},\\
\tilde{A}_{ij}&\rightarrow& 
\tilde{A}_{ij}-\frac{\tilde{A}}{3}\tilde{\gamma}_{ij},
\end{eqnarray}
as is commonly done in numerical relativity codes.  We do not,
however, enforce the Hamiltonian, momentum, or Gamma constraints,
\begin{eqnarray}  %DMSB (16)
\label{eq:hamcon}
   0 = \mathcal{H} &=&
                 \tilde\gamma^{ij}\tilde D_i\tilde D_j e^{\phi}
                 - {e^{\phi} \over 8}\tilde R  \\
                & & + {e^{5\phi}\over 8}\tilde A_{ij}\tilde A^{ij}
                    - {e^{5\phi}\over 12}K^2 + 2\pi e^{5\phi}\rho,
        \nonumber \\
\label{eq:momcon}%DMSB (16)
   0 = {\mathcal{M}}^i &=&
   \tilde D_j(e^{6\phi}\tilde A^{ji})- {2\over 3}e^{6\phi}\tilde D^i K
   - 8\pi e^{6\phi}S^i,\\
0 = \mathcal{G}^i &=& \tilde \Gamma^i + \tilde \gamma^{ij}_{~~,j},
\end{eqnarray}
so these serve as an independent check on the validity of our code.  
We do not add a Hamiltonian constraint damping term to the right hand side (RHS) 
of the $\phi$ evolution equation, but we add damping terms to the RHS of the
$\tilde{\Gamma}^i$, $\tilde{\gamma}_{ij}$, and $\tilde{A}_{ij}$ BSSN
evolution equations, following the prescription defined in
\cite{DSY04}.  These terms are zero analytically, and serve only to
stabilize evolutions.

\subsection{Hydrodynamic evolution}\label{sec:num_hydro}

The hydrodynamics equations are calculated using the HRSC scheme
described by \cite{DLSS}.  To recover the ``primitive variables''
$\rho_0$, $P$, and $v^i$ from the conserved set $\rho_*$,
$\tilde{\tau}$, and $\tilde{S}_i$, we perform the inversion as
specified by Eqs.~(57)--(62) of \cite{DLSS}.
%First, we rewrite the normalization of the 4-velocity,
%$u_{\mu} u^{\mu} = -1$, as
%\begin{equation}  %DLSS (57)
%w^2 = \rho_{\star}^2 + \gamma^{ij}\frac{\tilde{S}_i\tilde{S}_j}{h^2},\label{eq:unorm}
%\end{equation}
%where $w \equiv \alpha u^0 \rho_{\star}$.  Next, we write $h$ in terms
%of this and the conserved variables,
%\begin{equation} %DLSS (58)
%h = \frac{\Gamma w(\tilde{\tau}+\rho_{\star}) - (\Gamma-1)\rho_{\star}^2}
%    {\Gamma w^2-(\Gamma-1)\rho_{\star}^2} \  .
%\label{eq:heqn}
%\end{equation}
%Substituting Eq.~(\ref{eq:heqn}) into Eq.~(\ref{eq:unorm}) yields a 
%quartic
%equation for $(w-\rho_{\star})$ that we solve using a standard
%polynomial root finder, which combined with Eq.~(\ref{eq:heqn}) yields
%$h$.  The remaining primitive variables are calculated from the
%expressions
%\begin{eqnarray} %DLSS (59)-(62)
%u^0 & = & \frac{w}{\alpha \rho_{\star}} \\
%\rho_0 & = & \frac{\rho_{\star}^2}{\sqrt{\gamma} w} \\
%P & = & \frac{\Gamma-1}{\Gamma}\rho_0 (h-1) \\
%v^i & = & \frac{1}{u^0}\gamma^{ij}
%          \frac{\tilde{S}_j}{\rho_{\star}h} - \beta^i \ .
%\end{eqnarray}
As in \cite{FBEST}, our inversion algorithm occasionally finds
unphysical sets of conserved variables at points immediately adjacent
to the puncture and in our atmosphere, which do not allow for
solutions of the primitive variables.  As in that paper, we enforce
the following two conditions, which are both necessary and sufficient
to allow for a well-defined inversion everywhere, and result in smooth
hydrodynamic variable profiles in the BH interior after the puncture
has passed through a set of grid points:
\begin{eqnarray}  %If S^2 > tau(tau+2*rho), w > tau+rho and from DLSS
                  %(58), h<1.0
|\tilde{S}|^2\equiv\gamma^{ij}\tilde{S}_i\tilde{S}_j &<& 
\tilde{\tau}(\tilde{\tau}+2\rho_*),\\
\tilde{\tau}&>& 0.
\end{eqnarray}
When these conditions are not met we rescale $\tilde S^i$ so that its
new magnigude is $|\tilde{S}|^2=0.98
\tilde{\tau}(\tilde{\tau}+2\rho_*)$, and set
$\tilde{\tau}=10^{-18}\tilde{\tau}_{0;max}$, where
$\tilde{\tau}_{0;max}$ is the maximum value of $\tilde{\tau}$ present
in our initial data.

To stabilize our hydrodynamic scheme in regions where there is no
matter, we maintain a tenuous atmosphere on our grid, with a density
floor set to $10^{-10}$ of the maximum density on our grid at $t=0$.
The initial atmospheric pressure $P_{\rm atm}$ is set to the cold
polytropic value $P_{\rm atm} = \kappa \rho_{\rm atm}^{\Gamma}$, 
where $\kappa$ is the polytropic constant at $t=0$.  Throughout the
evolution, we impose limits on the atmospheric pressure to prevent
spurious heating and negative values of the internal energy
$\epsilon$.

\subsection{Boundary Conditions}\label{sec:num_bc}

We apply Sommerfeld outgoing wave boundary conditions to the entire
set of field and gauge variables ${\bf f}$
\begin{equation}
\label{wavelike} 
{\bf f}(r,t) = {r - \Delta r\over r}{\bf f}(r-\Delta r, t-\Delta T)
\end{equation}
on the outer boundary of our numerical grid.  Here $\Delta T$ is the
timestep and $\Delta r = \alpha e^{-2\phi}\Delta T$, where radii are
evaluated in physical (as opposed to fisheye) coordinates.  To enforce
these boundary conditions, we first transform $\beta^i$, $\phi$,
$\tilde{A}_{ij}$, and $\tilde{g}_{ij}$ into physical coordinates,
apply the Sommerfeld condition as specified above, and then transform
back to fisheye coordinates.  We do not transform $\tilde{\Gamma}^i$,
assuming that its value propagates outward independent of the radius.

Since we adopt the moving puncture method (as opposed to black hole
excision) there are no interior boundaries or boundary conditions.

\subsection{Diagnostics}\label{sec:num_diags}

To validate our calculations, we compute surface integrals for the
system ADM mass $M$, linear momentum $P_i$, and angular
momentum $J_i$, given by
\begin{eqnarray}
M&=&\frac{1}{2\pi}\oint \left(\frac{1}{8}\tilde{\Gamma}^i -
\tilde{\gamma}^{ij}\partial_j\psi \right)d\Sigma_i,\label{eq:madm}\\
P_i&=&\frac{1}{8\pi}\oint (K^j_i-\delta^j_i
K)d\Sigma_j,\label{eq:padm}\\
J_i&=&\frac{1}{8\pi}{\epsilon_{ij}}^k\oint x^j(K^m_k-\delta^m_k
K)d\Sigma_m,\label{eq:jadm}
\end{eqnarray}
where $\psi = e^{\phi}$ and $d\Sigma_i = (x^i/r) \psi^6 r^2\sin\theta
d\theta d\varphi$ for a spherical surface at fixed radius.  Note that the
above expressions are valid only if the spatial 3-metric $\gamma_{ij}$
approaches the Minkowski metric $\eta_{ij}$ at large $r$. Hence we
need to transform the variables from fisheye to physical coordinates
before integrating.  In addition, we monitor the following normalized
expressions for the Hamiltonian and momentum constraints
\begin{eqnarray}
   ||\mathcal{H}|| &\equiv& \int_{\mathcal{V}} (|\mathcal{H}|/N_{\text{HC}}) dV
 \label{eq:normH}  \\
   ||\mathcal{M}^i|| &\equiv& \int_{\mathcal{V}} (|\mathcal{M}^i|/N_{\text{MC}}) 
dV, \label{eq:normMi}
\end{eqnarray}
where
\begin{eqnarray}
   N_{\text{HC}} &=& \left( \left(2\pi\psi^5\rho\right)^2
   + \left(\tilde D^i\tilde D_i\psi\right)^2
   + \left({\psi\over 8}\tilde R\right)^2 \right. \\
   & & + \left. \left({\psi^5\over 8}\tilde A_{ij}\tilde 
A^{ij}\right)^2
      + \left({\psi^5\over 12}K^2\right)^2 \right)^{1/2},
\nonumber\\
  N_{\text{MC}} &=&
  \left(\sum_{i=1}^3\left[(8\pi S^i)^2
     + \left({2\over 3}\tilde D^iK\right)^2 \right.\right. \\
   & & + \left.\left.
     \left(\psi^{-6}\tilde D_j(\psi^6\tilde A^{ij})\right)^2
     \right]\right)^{1/2}.\nonumber
\end{eqnarray}

The apparent horizon (AH) of the BH is computed using the {\tt
ahfinderdirect} Cactus thorn~\cite{ahfinderdirect}.  This thorn
outputs the BH irreducible mass, which is related to the AH area
$\mathcal A$ as follows:
\begin{equation}
M_{\rm irr}=\sqrt{{\mathcal A}/16\pi}. \label{eq:mirr}
\end{equation}
With the AH surface computed by {\tt ahfinderdirect}, we can evaluate
diagnostic integrals in a region interior or exterior to the AH.  
%For example, we measure the baryon rest mass %$M_0=\int \rho_*
%dV$ inside the AH, $M_{\rm AH}(t)$.

\subsection{Gravitational radiation}\label{sec:num_gw}

To measure the gravitational wave (GW) emission from our binaries, we
use both gauge-invariant theory based on perturbations of a background
Schwarzschild spacetime derived by Zerilli \cite{Zerilli} and Moncrief
\cite{Moncrief}, hereafter referred to as the ``Z-M'' formalism, as
well as a technique that makes use of the Newman-Penrose Weyl scalar $\psi_4$.  In the
Z-M formulation, deformations of the spatial metric are
viewed as perturbations on a Schwarzschild background at large radii.
We decompose these perturbations into gauge-independent even and
odd-parity modes, denoted $\Psi^{lm}_{even}$ and $Q^{lm}_{\rm M}$ in
the notation of \cite{RTAN}, whose derivation is outlined below.
For convenience, we define the time-integral of the odd-parity mode
amplitude
\begin{equation}
\Psi^{lm}_{odd}\equiv -\int_{-\infty}^t Q^{lm}_{\rm M} dt'.\label{eq:mon_odd}
\end{equation}
In terms of these expressions, the complex gravitational wave strain
$H\equiv h_+-ih_{\times}$ is given by Eq.~(4.34) of \cite{RTAN},
\begin{equation}
H=\frac{1}{2r}\sum_{l,m}\sqrt{\frac{(l+2)!}{(l-2)!}}(\Psi^{lm}_{even}+i\Psi^{lm}_{odd})
_{-2}Y^{lm},\label{eq:moncrief}
\end{equation}
where $_{-2}Y^{lm}$ is the $s=-2$ spin-weighted spherical harmonic.
In the notation of Shibata and collaborators (see, e.g.,
\cite{SS05} and earlier papers), $R^{\rm E}_{lm}$ and
$R^{\rm O}_{lm}$ are related to our adopted notation as follows:
\begin{eqnarray}
R^{\rm E}_{lm}&\equiv& 
\frac{1}{\sqrt{2}r}\sqrt{\frac{(l+2)!}{(l-2)!}}\Psi^{lm}_{even},\\
R^{\rm O}_{lm}&\equiv& 
\frac{1}{\sqrt{2}r}\sqrt{\frac{(l+2)!}{(l-2)!}}\Psi^{lm}_{odd}.
\end{eqnarray}

In addition, we use the {\tt PsiKadelia} thorn to compute the complex
Weyl scalar $\psi_4$, which depends on the spatial metric and
extrinsic curvature.  The wave strains are given in terms of
$\psi_4$ by Eq.~(3.4) of \cite{RTAN},
\begin{equation}
H=-\int_{-\infty}^t \int_{-\infty}^{t'} \psi_4 dt' dt''. \label{eq:psi4}
\end{equation}
We also decompose $\psi_4$ into $s=-2$ spin-weighted spherical
harmonic modes.

In terms of $H$, the radiated energy, linear momentum, and z-component
of the angular momentum are calculated from Eqs.~(2.8), (2.11), and
(2.13) of \cite{RTAN}
\begin{eqnarray}
\frac{dE}{dt}&=&\lim_{r\rightarrow\infty}\frac{r^2}{16\pi}\oint
|\dot{H}|^2 d\Omega,\\
\frac{dP^i}{dt}&=&\lim_{r\rightarrow\infty}\frac{r^2}{16\pi}\oint
\frac{x^i}{r}|\dot{H}|^2 d\Omega,\\
\frac{dJ_z}{dt}&=&-\lim_{r\rightarrow\infty}\frac{r^2}{32\pi}\oint
H^* \partial_\phi H d\Omega.
\end{eqnarray}

In practice, radiative losses are computed as a sum over modes 
(up to and including $l=4$),
following expressions equivalent to Eqs.~(4.41), (4.43), and (4.47)
of \cite{RTAN} for the Z-M case, and  Eqs.~(3.6), (3.8), (3.14),
and (3.24) for the $\psi_4$ case. 

To compute the gravitational wave energy spectra, we use the same techniques as
\cite{STU2}, determining the energy loss per unit frequency from
their Eq.~(8),
\begin{equation}
\frac{dE}{df}=\frac{\pi}{8}\sum_{l,m}f^2 \frac{(l+2)!}{(l-2)!}
|\tilde{\Psi}_{lm}|^2,
\end{equation}
where $|\tilde{\Psi}_{lm}|\equiv |\tilde{\Psi}^{lm}_{even}|^2
   +|\tilde{\Psi}^{lm}_{odd}|^2$, and $\tilde{\Psi}\equiv \int
   e^{2\pi i f t}\Psi(t)dt$.  We also define the
effective gravitational wave amplitude, taking the $z\to0$ limit of
Eq.~(5.1) from \cite{FH}:
\begin{equation}
h_{\rm eff}(f)\equiv \frac{\sqrt{2}}{\pi 
r}\sqrt{\frac{dE}{df}}.\label{eq:heff}
\end{equation}
Notice that this expression differs by a proportionality constant
from that found in \cite{Shi05}.  This is due to different assumptions
regarding the alignment of the binary and viewing angle relative to
the detectors.  The constant here is specified to reflect the RMS
average of signal amplitudes over all possible orientations of merging
binaries and the detector.

\section{Numerical Calculations of BHNS merger}\label{sec:runs}

To begin our survey of the BHNS merger parameter space, we focus on
those configurations likely to undergo mass loss and disk
formation, i.e., systems with binary mass ratios comparable to unity.
To ensure that our results are physically relevant, we
consider cases with appropriately large NS compactions,
noting that lowering the compactions would certainly
increase the mass of a potential disk.

All our calculations begin with initial configurations taken from
\cite{TBFS07a}, which consist of irrotational NSs orbiting
approximately nonspinning BHs.  We investigate two different cases for
the NS compaction, distinguished by nondimensional mass
$\bar{M}_0\equiv M_0/\kappa^{1/2}=0.15$ and $0.1$, where $M_0$
is the rest (baryon) mass of the NS, and $\kappa$ is the polytropic constant.
We refer to these as the high and low-compaction NS cases.  These
configurations yield compactions $\mathcal{C}=M_{\rm NS}/R_{\rm NS}$ of
0.145 and 0.0879, respectively, where $M_{\rm NS}$ is the ADM mass and
$R_{\rm NS}$ is the radius of the NS in isolation.  Setting $\kappa$
fixes the NS mass for a given compaction and polytropic index
\cite{CookShapTeuk}.
Choosing the high (low) compaction NS to have a rest mass of
$1.4M_\odot$, we find the ADM mass for the isolated NS to be
$1.30M_\odot$ ($1.34M_\odot$), with an isotropic radius of $11.24~{\rm km}$
($20.46~{\rm km}$), and circumferential (Schwarzschild) radius of
$13.24~{\rm km}$ and ($22.49~{\rm km}$).

We consider data from four different independent sequences in
\cite{TBFS07a}, consisting of the high-compaction NS in binaries with
mass ratios $q\equiv M_{\rm irr}/M_{NS}=3$, $2$, and
$1$, as described in their Table~IV, along with the low-compaction NS
for a binary mass ratio $q=2$, from their Table~V. Here, $M_{\rm irr}$
is the BH irreducible mass.  These sequences are
denoted ``A''-``D'', respectively, and are summarized in
Table~\ref{table:initconf}.

For sequence A, we consider three separate initial binary separations
(denoted A1, A2, and A3) to determine how the initial separation 
affects the results of
the simulations.  For case A1, we performed a series of simulations
designed to study the effect of the numerical grid parameters on our
results.  In cases A1-lo, A1-med, and A1-hi, we varied only the grid
resolution, keeping the outer boundary at a fixed location.  Case
A1-farbc has the same grid resolution as A1-med, but the outer
boundary is farther away.

Performing multiple runs with the same quasiequilibrium binary
parameters (sequence A) enables us to gauge the resources necessary to
perform accurate evolutions and validate our physical results.  For
example, we may compare the location of the tidal disruption point against
quasiequilibrium estimates of \cite{TBFS07b}.  We also investigate how
the disk mass varies with regard to our grid parameters, noting that
this had a significant effect on the results found in \cite{SU1,SU2}.
Finally, performing multiple runs allows us to check the robustness of
the gravitational wave signals.

For convenience, we identify the approximate time at which the
low-density regions of the deformed NS first fall into the BH horizon as the
moment of ``first contact'', $t_{\rm FC}$.  

\begin{table*}
\caption{Summary of our initial configurations, which are taken from
  the results of \protect\cite{TBFS07a}.  Here, the mass ratio
  $q\equiv M_{\rm irr}/M_{NS}$, $\bar{M}_0\equiv
  M_0/\kappa^{1/2}$ is the nondimensional rest (baryon) mass,
  $M$ is the total ADM mass for the binary
  system (assuming that $M_0 = 1.4M_{\odot}$), $a_0$ the initial binary coordinate separation, $J$ the
  initial angular momentum of the system, $\Omega$ the orbital
  frequency, and $t_{\rm FC}$ the time of first contact.
  Finally, $M\Omega_{\rm CTS}$ is the frequency of tidal disruption,
  as derived from our initial data \cite{TBFS07a}, and $M\Omega_{\rm
  num}$ is the frequency at which our numerically derived waveform
  spectrum deviates from the restricted post-Newtonian value by $25\%$ (see
  Section~\ref{sec:gw_emission}).
  }
\begin{tabular}{|c cc c c c c c c c|}
  \hline
  {\bf Case} & $q$ & $\bar{M}_0$ & $M/M_\odot$ & $a_0/M$ &
  $J/M^2$ & $M\Omega$ & $t_{\rm FC}/M$ & $M\Omega_{\rm CTS}$ &
  $M\Omega_{\rm num}$ \\
  \hline
A1 & 3.0 & 0.15 & 5.15 & 5.41 & 0.629 & 0.0628 & 50 & 0.0728 & -- \\
%Old notation:A5,A1,A3,A6, respectively 
A2 & 3.0 & 0.15 & 5.16 & 6.49 & 0.663 & 0.0435 & 105 & 0.0728 & -- \\
%Old notation: A2
A3 & 3.0 & 0.15 & 5.17 & 8.81 & 0.698 & 0.0329 & 215 & 0.0728 & 0.0789 \\
%Old notation: A4
B & 2.0 & 0.15 & 3.86 & 7.17 & 0.774 & 0.0499 & 60 & 0.0550 & 0.0656\\
C & 1.0 & 0.15 & 2.57 & 8.61 & 0.936 & 0.0343 & 91 & 0.0382 & 0.0509\\
D & 2.0 & 0.1 & 3.98 & 11.56 & 0.909 & 0.0228 & 155 & 0.0255 &  0.0395\\
%Old notation: B1, C2, D1
\hline
\end{tabular}
\label{table:initconf}
\end{table*}

\begin{table*}
\caption{Summary of the grid configurations and fisheye parameters 
used for our runs.}
\begin{tabular}{|c cc c c cc|}
  \hline
  Run & $\Delta x_{\rm int}/M$ & $\Delta x_{\rm ext}/M$ & Grid Size & $r_{\rm out}/M$ &
  $R_1/M$ & $s_1/M$ \\
\hline
A1-hi & 0.05 & 0.4 & $540^2 \times 270$ & 39.5 & 9.7 & 1.6 \\
A1-med & 0.061 & 0.49 & $440^2 \times 220$ & 39.5 & 9.7 & 1.6 \\
A1-lo & 0.081 & 0.65 & $332^2 \times 166$ & 39.5 & 9.7 & 1.6 \\
A1-farbc & 0.061 & 0.49 & $604^2 \times 302$ & 79.6 & 9.7 & 1.6 \\
%Old notation:A5,A1,A3,A6, respectively 
A2 & 0.061 &  0.49 & $532^2 \times 266$ & 39.7 & 12.9 & 1.6 \\
%Old notation: A2
A3 &  0.061 & 0.49 & $516^2 \times 258$ & 40.0 & 12.3 & 1.6 \\
%Old notation: A4
B &  0.081 & 0.65 & $404^2 \times 202$ & 50.2 & 11.6 & 2.2\\
C & 0.05 & 0.4 & $610^2 \times 305$ & 40.0 & 11.7 & 1.6 \\
D &  0.061 & 0.49 & $566^2 \times 283$ & 40.1 & 15.8 &  1.6 \\
% old notation: B1,C2,D1, respectively
\hline
\end{tabular}
\label{table:grid}
\end{table*}

\subsection{BHNS binaries with mass ratio 3:1}
\label{sec:caseA}
To begin our discussion, we consider results from evolutions of
sequence~A, the high compaction NS case with $\bar{M}_0=0.15$ and
$q=3$.

\begin{figure*}
\vspace{-4mm}
\begin{center}
\epsfxsize=3.5in
\leavevmode
\epsffile{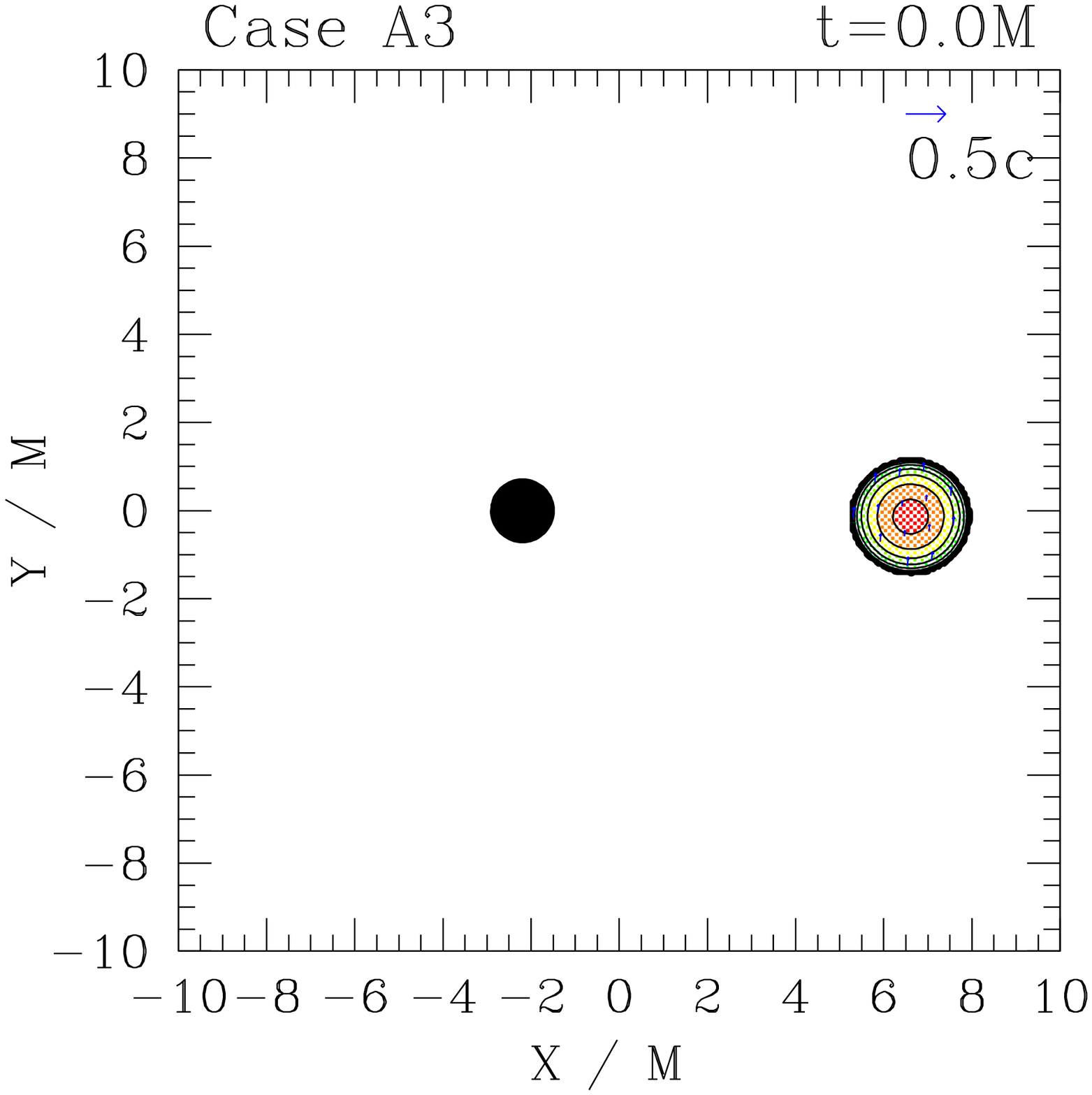} %2
\epsfxsize=3.5in
\leavevmode
\epsffile{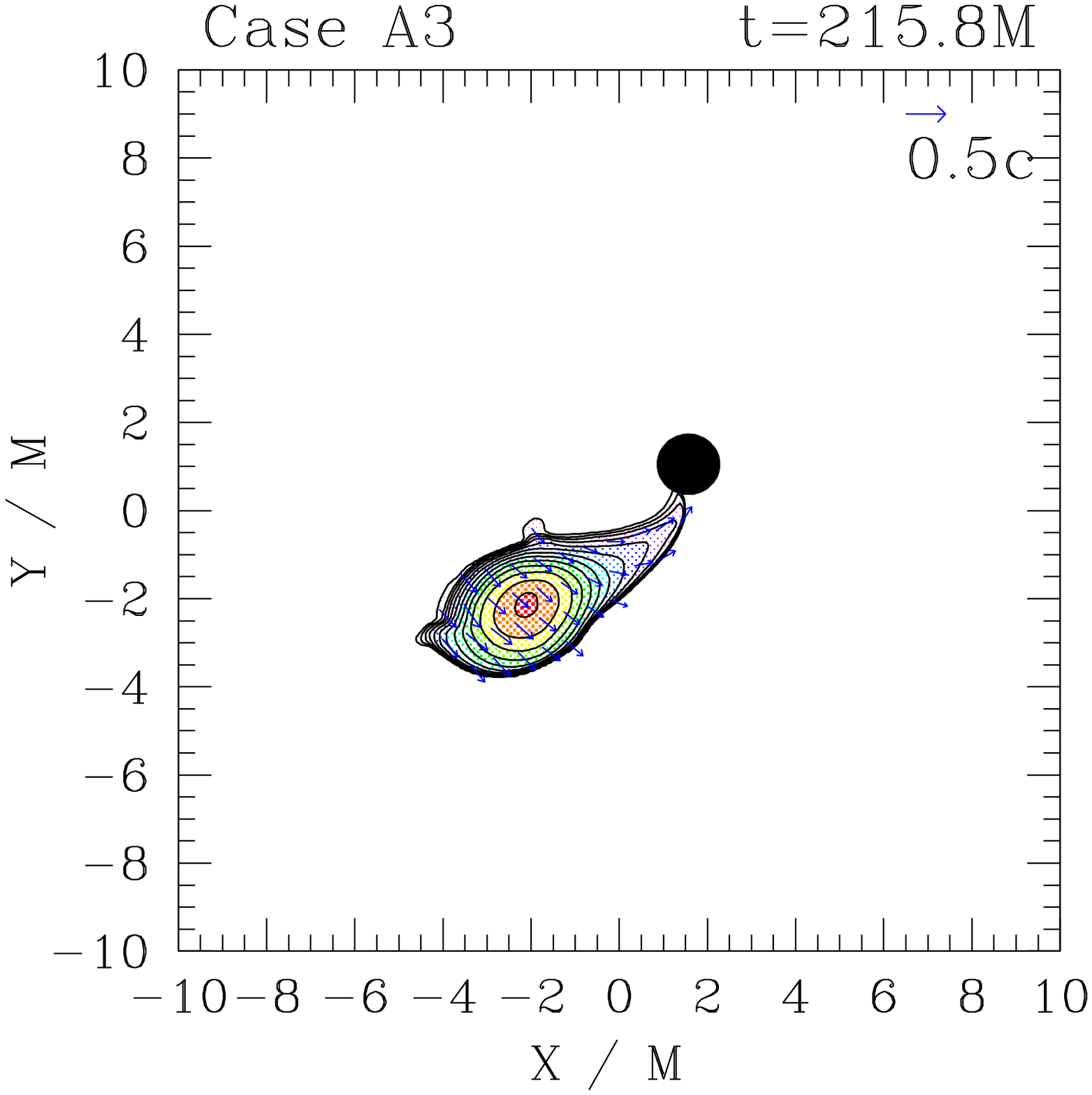}\\ %342
\epsfxsize=3.5in
\leavevmode
\epsffile{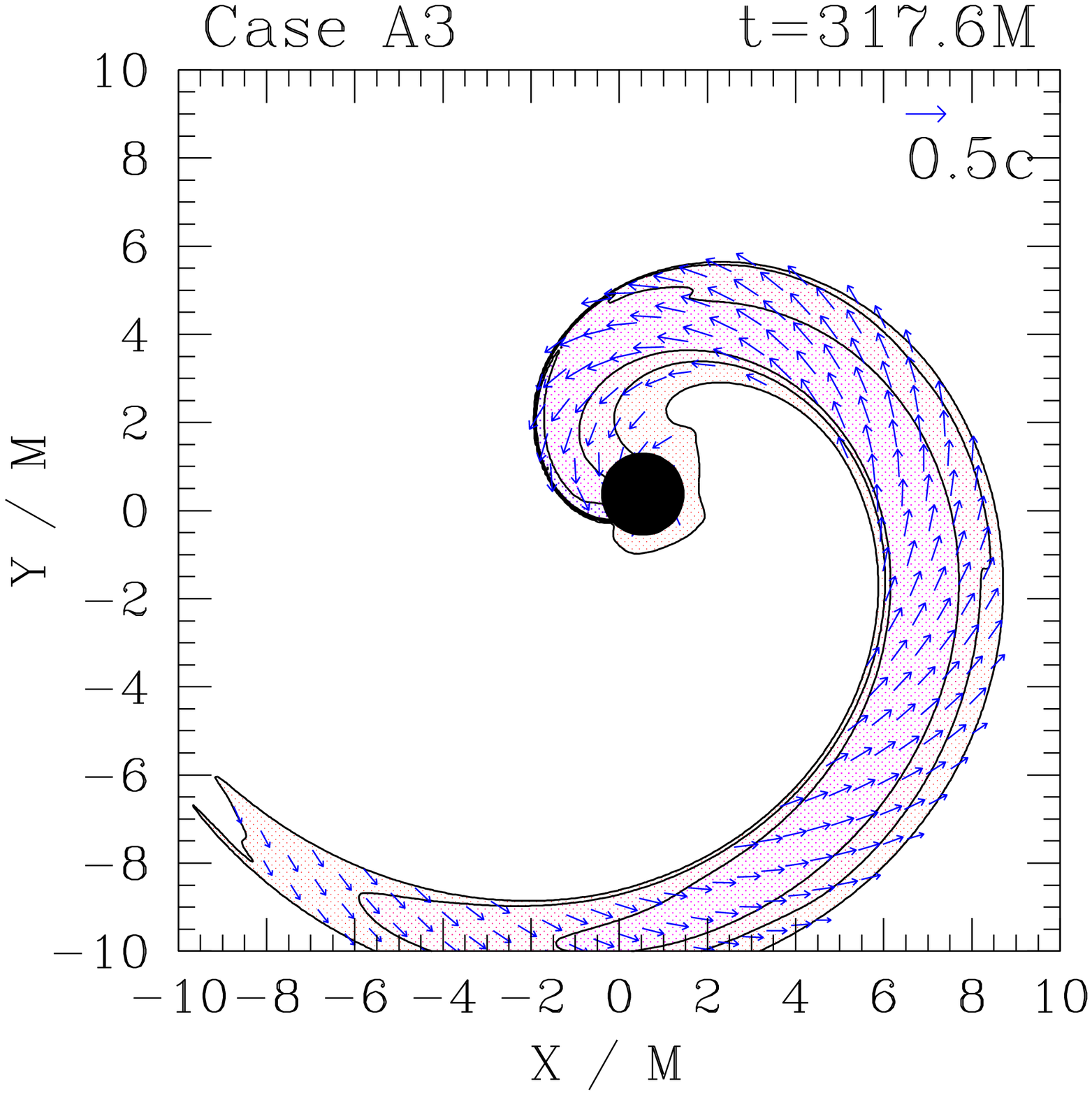} %500
\epsfxsize=3.5in
\leavevmode
\epsffile{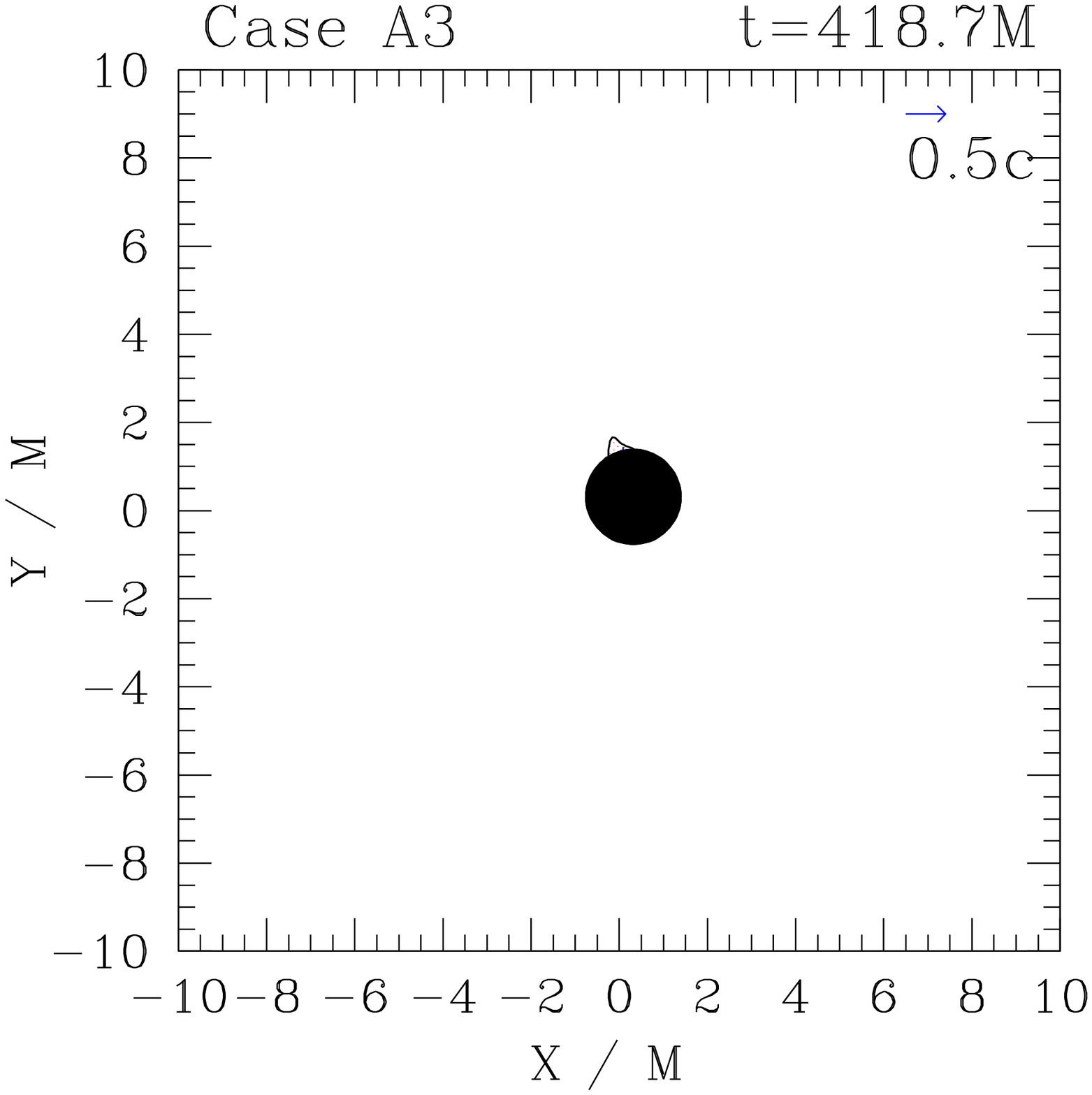} %660
\caption{Snapshots of density and velocity profiles at selected times
  for run A3, with binary mass ratio $q=3$.  The contours represent
  the density in the orbital plane, plotted logarithmically with four
  contours per decade, with greyscaling added for clarity.  Arrows
  represent the velocity field in the orbital plane.  The minimum
  contour value in each frame is $\kappa \rho_0$(min)$=10^{-4}$,
  or $\rho_0$(min)$=7\times 10^{11} (1.4M_\odot/M_0)^2$g
  cm$^{-3}$.  The maximum initial NS density is $\kappa \rho_0 = 0.13$.
We specify the black hole AH interior in each snapshot with a
  filled black circle.  In cgs units, the total ADM mass for this case is
  $M=3\times 10^{-5}(M_0/1.4M_\odot)$
  s$=8(M_0/1.4M_\odot)$km.}
\label{fig:xy_caseA3}
\end{center}
\end{figure*}

In Fig.~\ref{fig:xy_caseA3}, we
plot density contours with overlaid 3-velocity vectors in the equatorial
plane for our large initial separation case (A3).  The orbital
direction is counter-clockwise in these snapshots.  The upper left
panel shows the initial configuration.  We see the onset of
accretion after approximately 1.75 orbits ($t\approx215M$), with
matter flowing in a narrow stream through the inner Lagrange point and
into the BH.  The accretion flow then accelerates as the NS is
consumed by the BH.  Later, at $t\approx 290M$, we see the beginnings
of mass loss outward into a disk through the outer Lagrange point, but
the mass stream is quite tenuous.  At the end of our simulation
($t=418.7 M$), no more than 3\% of the total rest mass of the NS
exists outside the AH.  The remaining matter is gravitationally bound
to the BH.  However, not all of this matter will form what would
typically be referred to as a disk, i.e., a quasi-stationary torus that
evolves on secular rather than dynamical timescales.  Instead, some of
the exterior mass at the end of our calculations will be accreted on
relatively short timescales, until the remaining 
fraction achieves rotationally-supported quasi-equilibrium.  Thus,
these estimates should be taken as upper limits on true ``disk
masses'' for a particular set of initial data and grid parameters.

In case~A of~\cite{SU1}, SU perform a fully GR simulation of a BHNS
merger with a synchronized NS and a mass ratio of $q=2.47$.  They find
the fraction of the initial rest mass of the NS in the final disk to
be $\approx 19-28\%$ of the initial rest mass of the NS.  We find
this disk mass fraction to be $\lesssim 3\%$ for our $q=3$,
irrotational NS model~A3, which suggests that more slowly spinning NSs
feed less mass into a disk.  In case~A of~\cite{ST07}, ST simulate an
irrotational, $q=3.06$ BHNS binary and obtain a disk with a rest mass
that is $\approx 6.6\%$ of the initial NS rest mass.  While this
result is consistent with our observation that lower NS spin
suppresses the mass of the disk, ST still find a significantly larger
disk than we do.

\begin{figure}
\epsfxsize=3.4in
\leavevmode
\epsffile{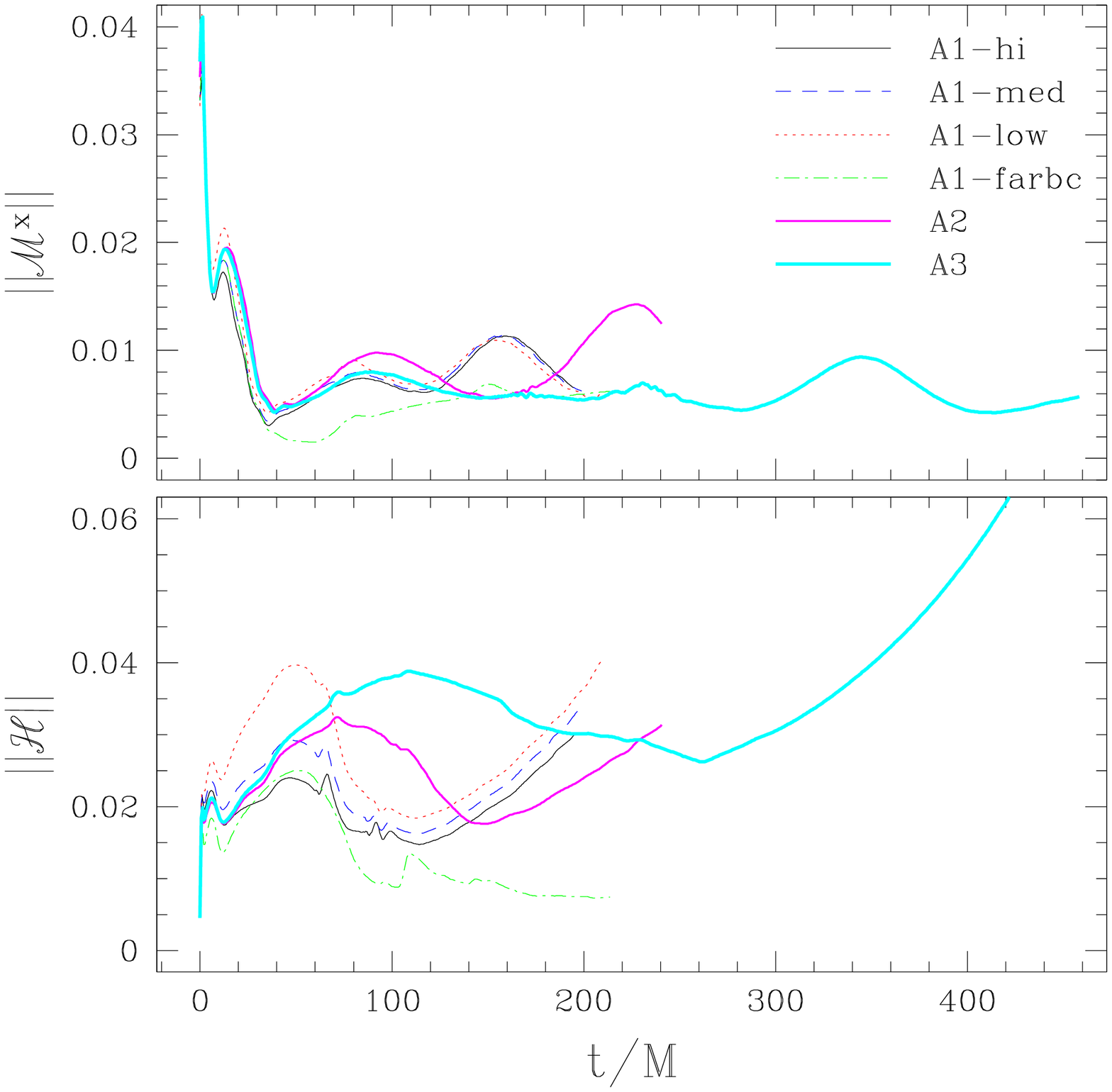}
\caption{Normalized violation of the $x$-component of the momentum
 constraint $||{\cal M}^x||$ (top panel) and the Hamiltonian constraint
$||\cal H||$ (bottom panel); see
 Eqs.~(\ref{eq:normH}) and~(\ref{eq:normMi})}.
\label{fig:cons_runs_a}
\end{figure}

To check the validity of our simulations, we monitor the normalized
Hamiltonian and momentum constraint violations using
Eqs.~(\ref{eq:normH}) and (\ref{eq:normMi}), respectively.  We show
the results from sequence~A in Fig.~\ref{fig:cons_runs_a}.
In all cases, the Hamiltonian and momentum constraint violations are
$\sim 3\%$ throughout the evolution.  We find that at early times, the
Hamiltonian constraint violation decreases as resolution is increased,
as expected. However, at late times the Hamiltonian constraint
violation steadily increases, and its magnitude becomes insensitive to
resolution (see lines corresponding to cases A1-hi, A1-med and A1-low
in Fig.~\ref{fig:cons_runs_a}), suggesting that the error is dominated
by reflection from the outer boundaries.  Indeed, we find that the
Hamiltonian constraint violation is significantly smaller when we move
the outer boundaries outward (see case A1-farbc in Fig.~\ref{fig:cons_runs_a}).

\begin{figure}
\epsfxsize=3.4in
\leavevmode
\epsffile{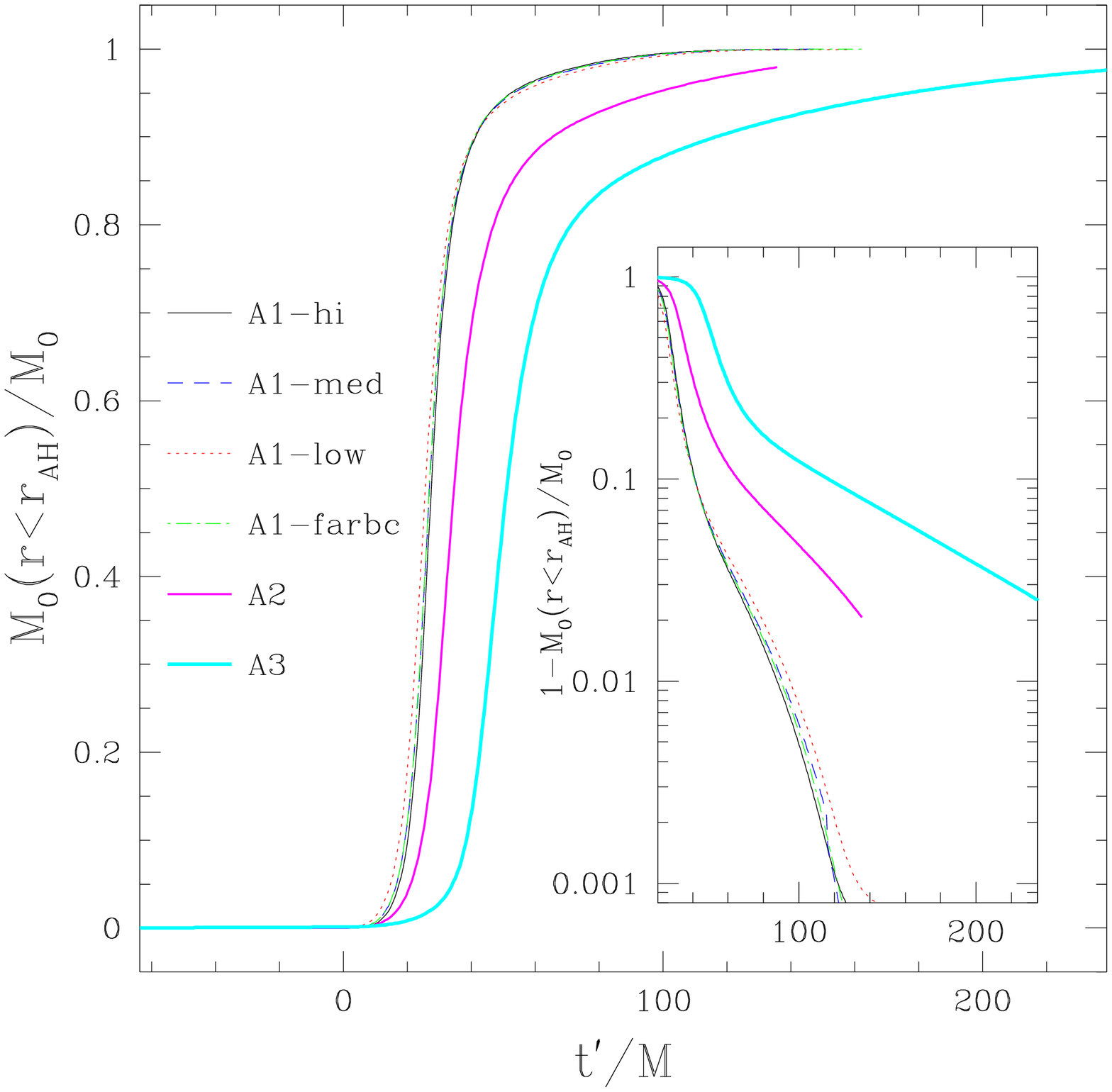}
\caption{Rest (baryon) mass fraction located inside the BH apparent horizon
  versus time for sequence~A. Note that time is shifted ($t'\equiv
  t-t_{\rm FC}$) so that the first contact 
  occurs at $t'=0$.  The inset shows the rest mass fraction outside
  the BH horizon at late times.}
\label{fig:mass_runs_a}
\end{figure}

Next, we investigate how our results depend on both the initial binary
separation and the numerical resolution.  In
Fig.~\ref{fig:mass_runs_a}, we plot the fraction of the rest (baryon) mass
inside the apparent horizon, $f_{\rm in} \equiv M_0(r<r_{\rm
AH})/M_0$, as a function of time for sequence A.  We find that
$f_{\rm in}$ depends only very weakly on the resolution or the location of
the outer boundaries.  We do, however,
find stronger dependence on the initial binary separation (column $a_0/M$
in Table \ref{table:initconf}).   This dependence
may be due to the zero radial infall speed in our initial data, which
results in a slightly eccentric orbit that increases the radial infall
speed as the binary approaches the ISCO, affecting the tidal disruption
and the disk formation (compare \cite{Mil04,biw06,PfeBKLLS07,HusHGSB07}).
This effect may be compounded by the fact that our initial configuration
is very close to the ISCO.  Alternatively, this dependence may be caused by
growing numerical error during the simulation. Despite these
small uncertainties, we find $f_{\rm in} \gtrsim$ 97\% at the end of
our simulations in all our cases, suggesting that there is no
appreciable disk left behind after merger.

\begin{figure}
\epsfxsize=3.4in
\leavevmode
\epsffile{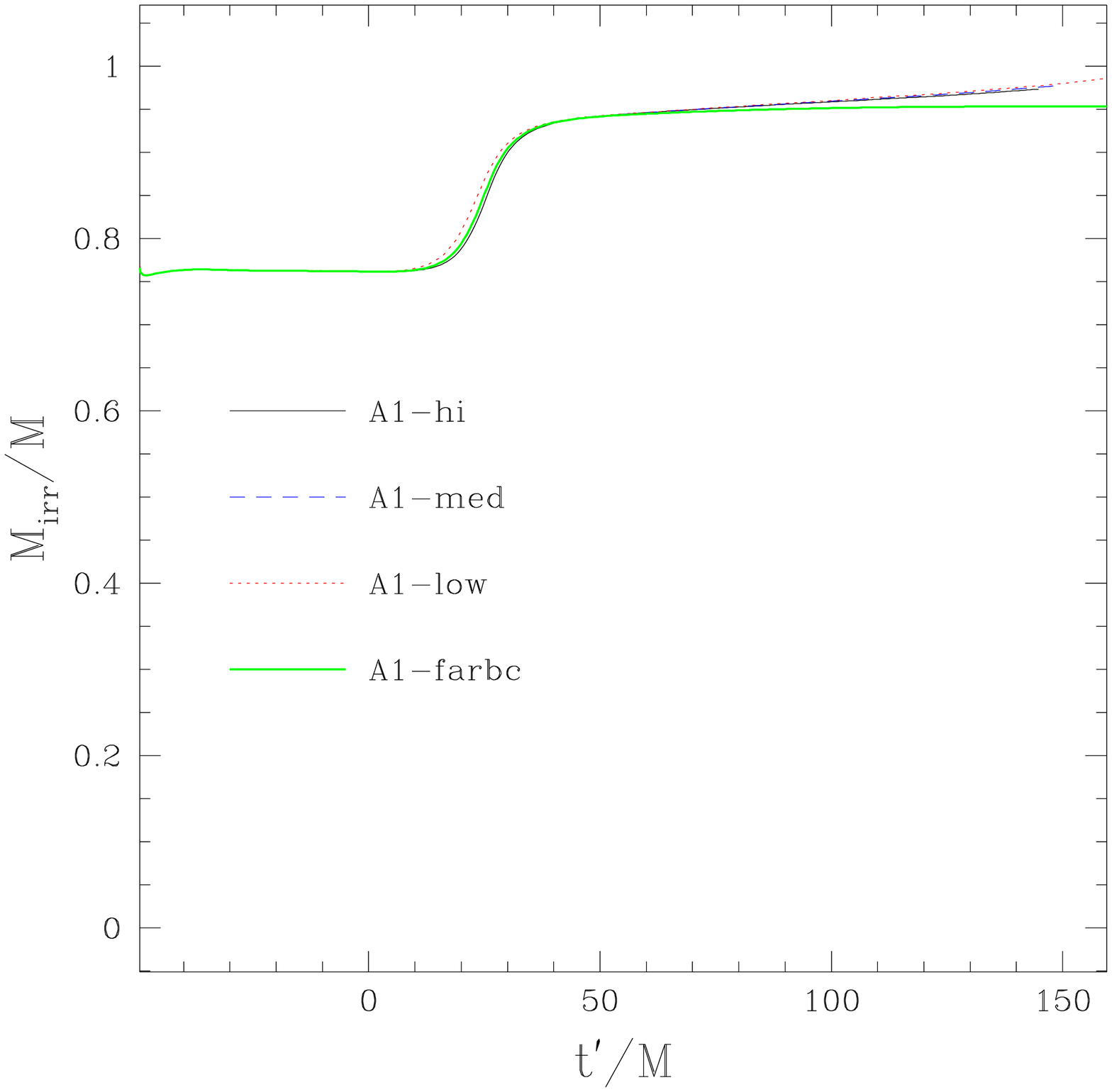}
\caption{Irreducible mass of the BH for sequence A1.  Here $M$ is the
  total ADM mass.}
\label{fig:mirr_runs_a}
\end{figure}

Figure~\ref{fig:mirr_runs_a} shows the evolution of the BH irreducible mass,
$M_{\rm irr}$, for the sequence~A1 simulations. We see that
$M_{\rm irr}$ increases as the NS matter is accreted, as expected.
At late times, $M_{\rm irr}$ approaches an asymptotic value when most
of the matter has fallen into the BH. However, reflection from the outer
boundaries causes $M_{\rm irr}$ to slowly increase at very late times
for runs A1-hi, A1-med and A1-low. This spurious effect is
significantly reduced when the outer boundaries are moved to a larger
radius (run A1-farbc), and $M_{\rm irr} \rightarrow 0.954M$ at late
times.

\subsection{The effect of the binary mass ratio and NS compaction}

\begin{figure*}
\vspace{-4mm}
\begin{center}
\epsfxsize=2.15in
\leavevmode
\epsffile{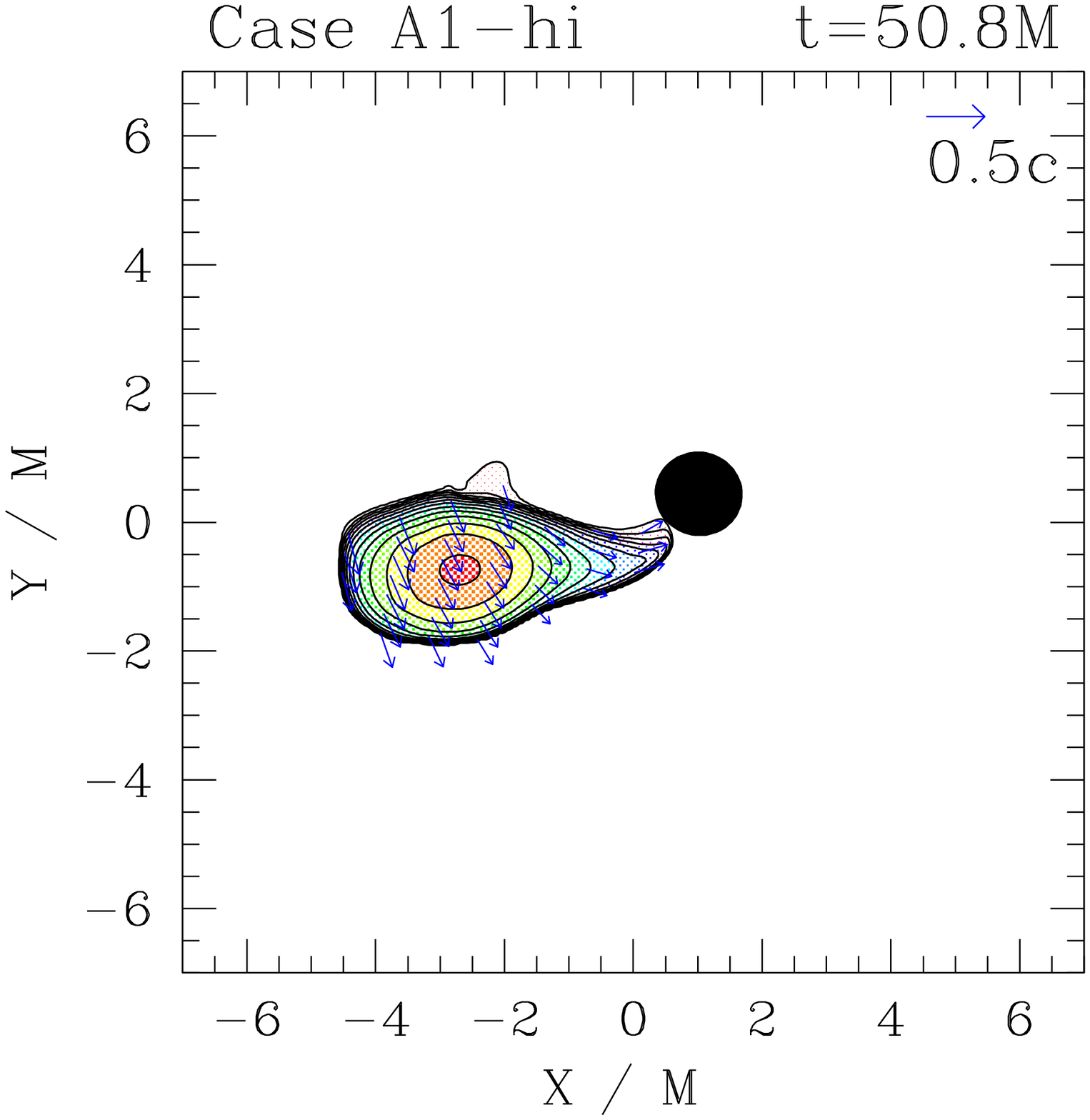}
\epsfxsize=2.15in
\leavevmode
\epsffile{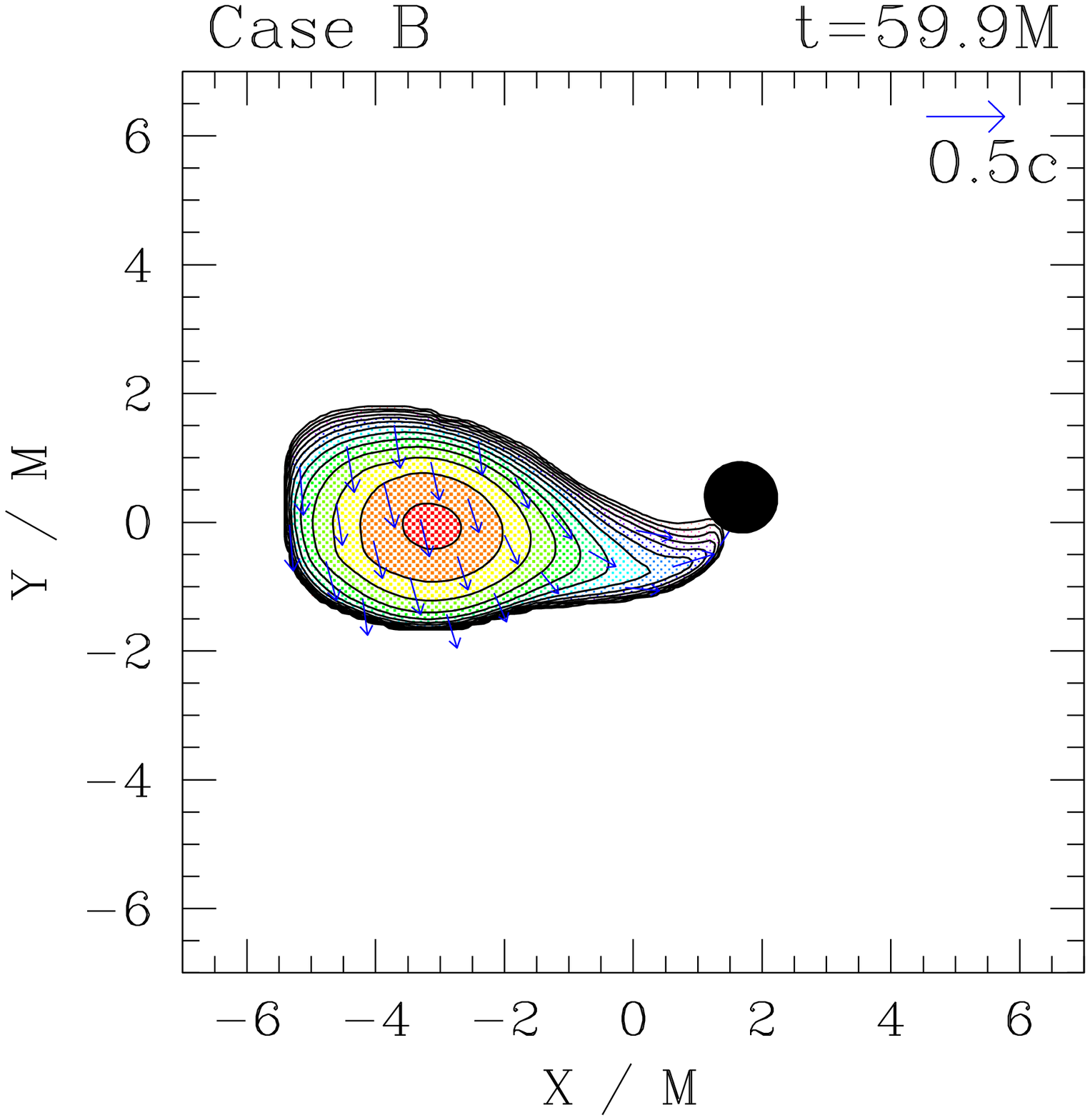} %70
\epsfxsize=2.15in
\leavevmode
\epsffile{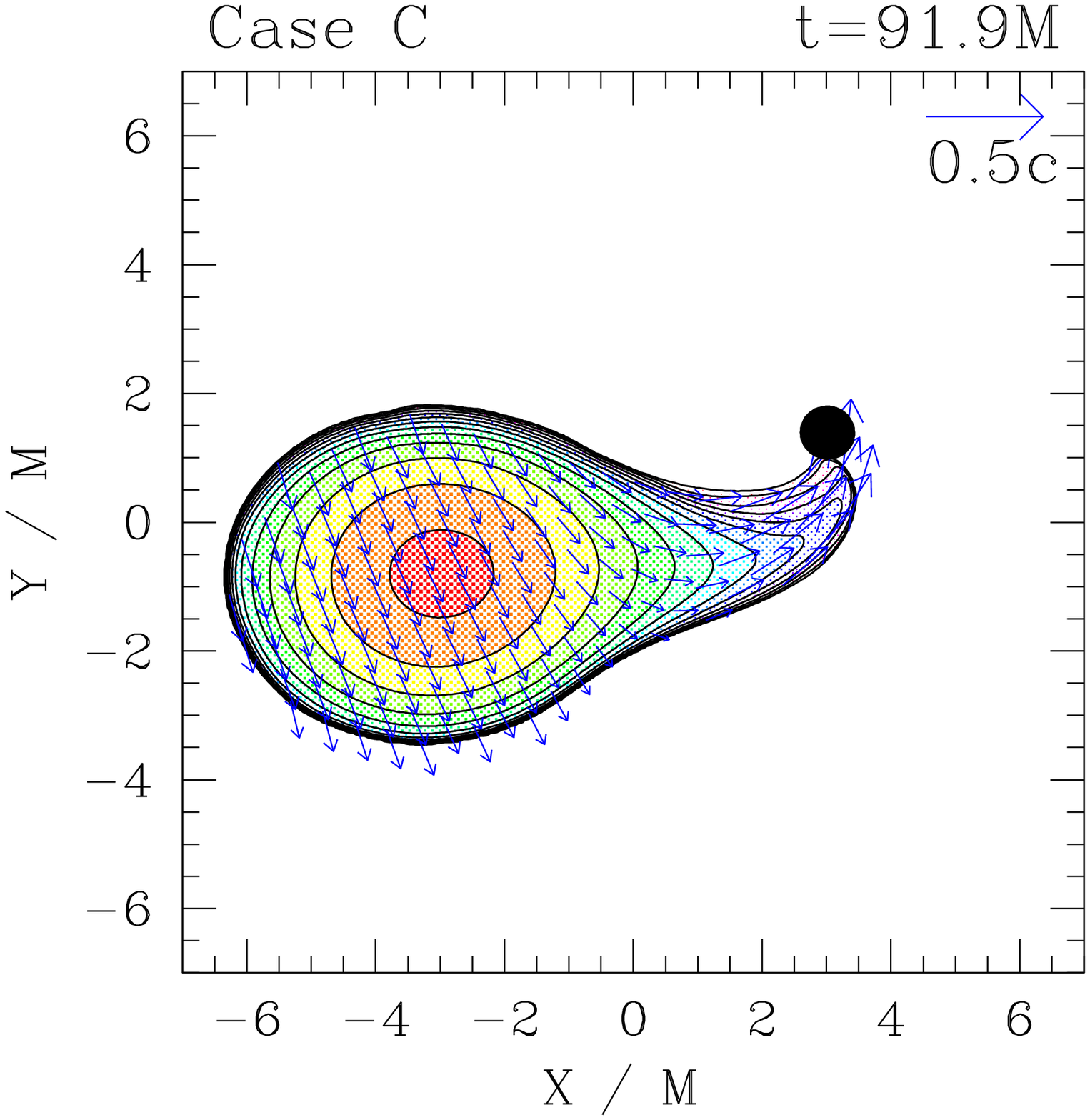}\\ %150
\epsfxsize=2.15in
\leavevmode
\epsffile{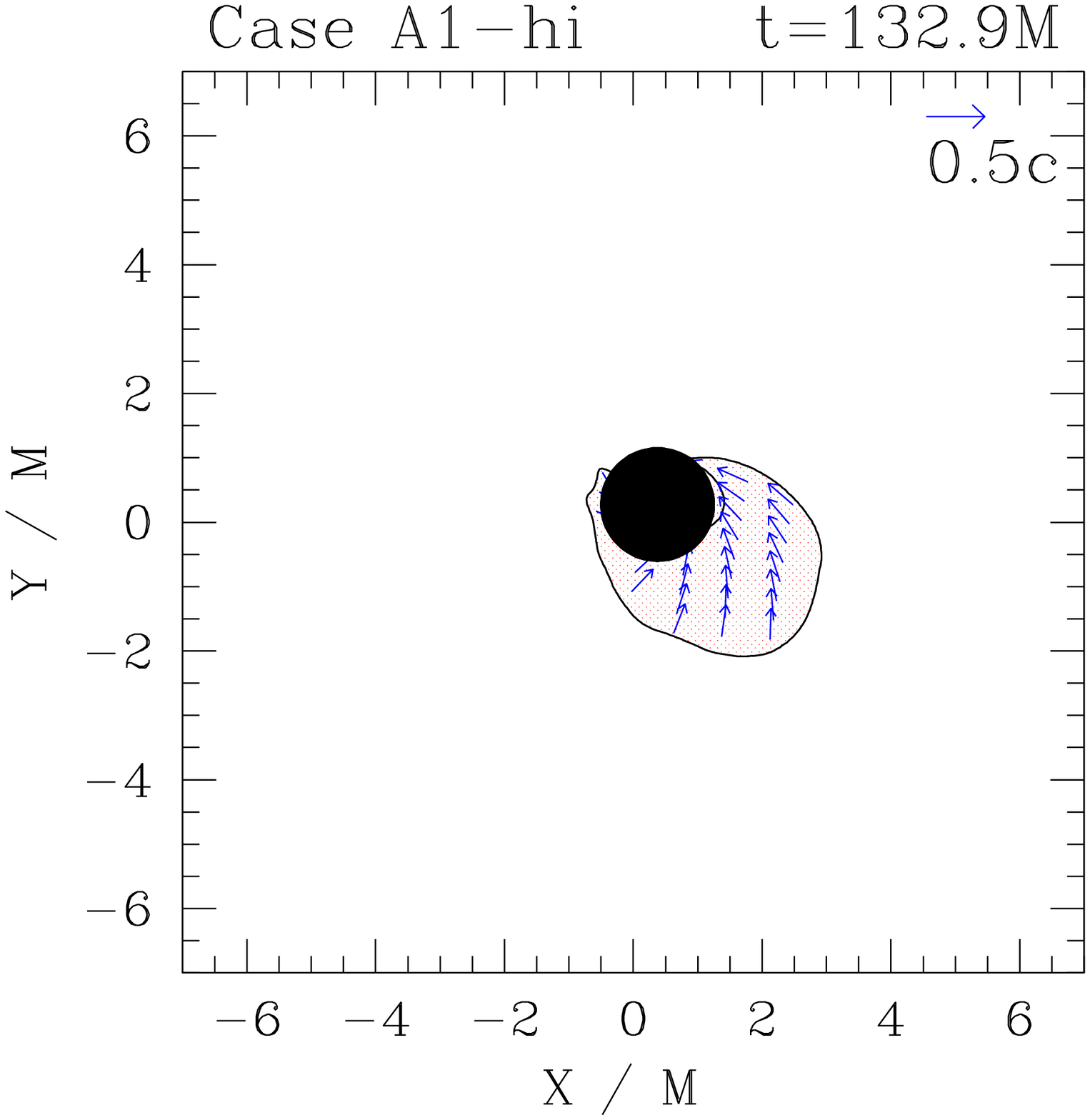} %456
\epsfxsize=2.15in
\leavevmode
\epsffile{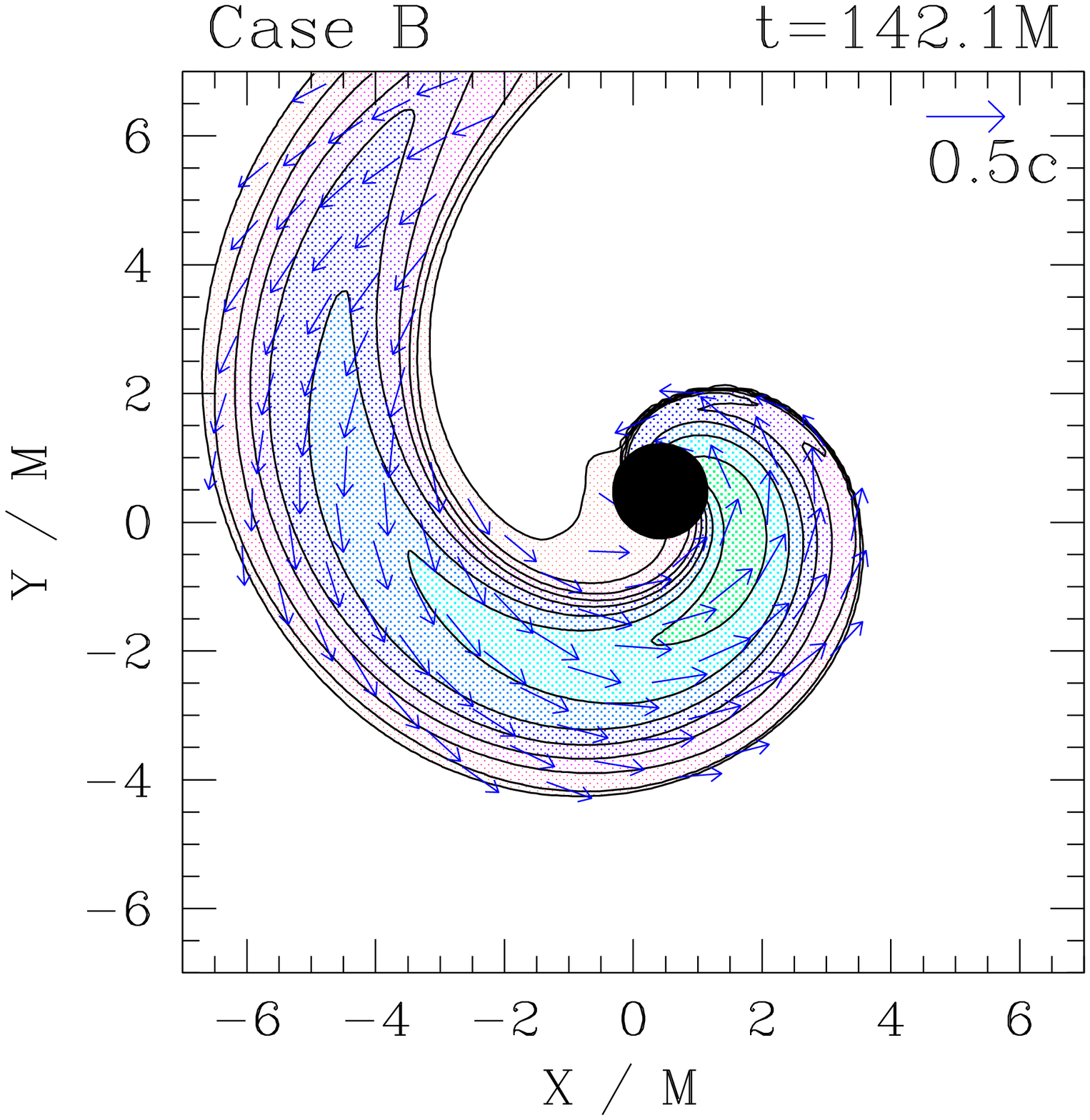} %158
\epsfxsize=2.15in
\leavevmode
\epsffile{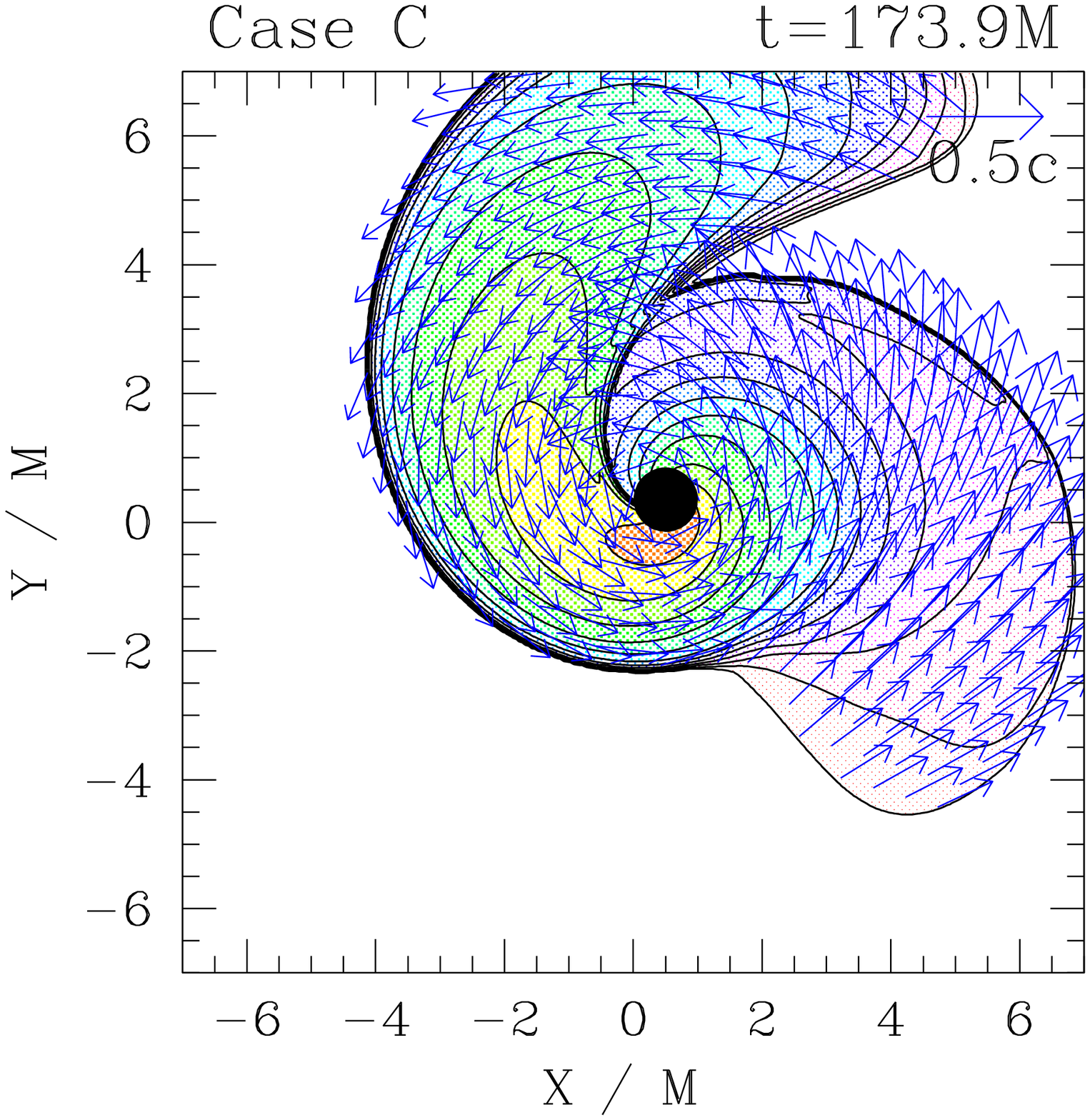} %270
\caption{Snapshots from runs A1-hi, B, and C, compared at first
   contact (upper panels) and at a moment in time $\Delta t\approx75M$
   later (lower panels).  Contours are defined as in
   Fig.~\ref{fig:xy_caseA3}.}
\label{fig:xy_seqABC}
\end{center}
\end{figure*}

Fig.~\ref{fig:xy_seqABC} demonstrates the variation in dynamics due to
a change in binary mass ratio, keeping the NS companion fixed.  All
three cases evolve approximately 0.5 orbits before the tidally
disrupted NS first touches the apparent horizon (i.e, the time of
``first contact'' -- the top 3 plots of Fig.~\ref{fig:xy_seqABC}).  At
this point in time, the general morphology of the systems are the
same, regardless of mass ratio: a funnel-shaped NS, with matter
flowing through the narrow end of the funnel into the BH.  After the
time of ``first contact'' ($t'\equiv t-t_{\rm FC}>0$), the dynamics of
the system depend most sensitively on the initial mass, and hence
size, of the black hole in comparison to the size of the NS.  Defining
$\chi^{\text{funnel}}_{\angle}$ as the (coordinate) angular extent of
the accretion funnel (with the density cutoff as defined in
Fig.~\ref{fig:xy_caseA3}) around the AH on the equatorial plane, we
find that as $\chi^{\text{funnel}}_{\angle}$ increases to $180^{\circ}$,
the accretion rate slows.

\begin{figure}
\epsfxsize=3.4in
\leavevmode
\epsffile{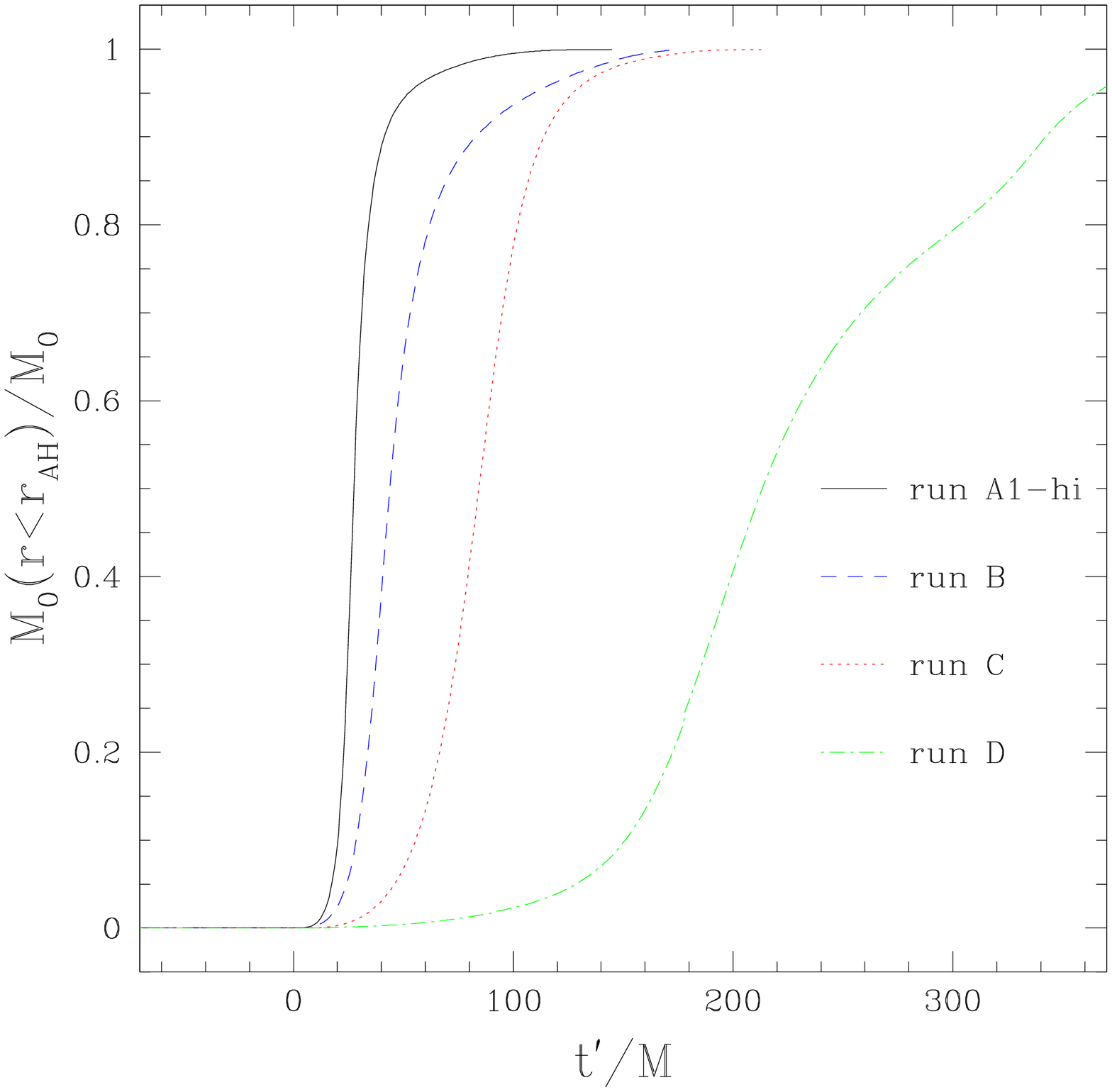}
\caption{Rest (baryon) mass located inside the BH apparent horizon versus
   $t'/M$ for runs A1-hi, B, C, D. }
\label{fig:massloss_seqABC}
\end{figure}

For example, in the A cases ($q=3$), the BH is sufficiently large so
that $\chi^{\text{funnel}}_{\angle}<180^{\circ}$ throughout much of
the simulation. We see that the NS accretion rate does not 
slow down until $\approx 90\%$ of the rest mass falls into the BH
(Fig.~\ref{fig:massloss_seqABC}).  After the densest part of the NS
falls into the BH ($t'\gtrsim30M$), the remaining NS funnel curls and
expands into a long, low-density tail while accretion
continues.  During the tail phase, the accretion rate slows until
$\approx99\%$ of the NS rest mass is inside the BH
(see Figs.~\ref{fig:xy_seqABC} and \ref{fig:massloss_seqABC}).

In case B ($q=2$), the BH is $33\%$ smaller and
$\chi^{\text{funnel}}_{\angle}$ reaches $\approx 180^{\circ}$ at
$t'\approx40M$, at about the time when the densest part of the NS core
falls into the BH.  The remaining material in the funnel begins to
evolve into a tail around this time, when only $\approx 55\%$ of the NS
rest mass has fallen into the BH, and the accretion rate decreases
considerably. Eventually the tail (Fig.~\ref{fig:xy_seqABC}) is
swallowed by the BH, but the accretion time scale is about twice as
long as the $q=3$ case (Fig.~\ref{fig:massloss_seqABC}).

At $t'\approx30M$ in case~C ($q=1$), a low-density region of matter
develops ahead of the higher-density funnel region as
$\chi^{\text{funnel}}_{\angle}$ becomes larger than
$180^{\circ}$.  As the majority of the NS material passes through
the high-density funnel, this low-density overshoot grows in size
away from the horizon while wrapping quickly around the BH, and finally
smashing into the high-density funnel region
(Fig.~\ref{fig:xy_seqABC}) before falling into the BH.  Once the
low-density overshoot has been accreted ($t'\approx125M$), only about
$15\%$ of the NS rest mass remains outside the BH.  The
remaining NS matter is accreted at about the same rate as in the $q=3$
case. 

\begin{figure*}
\vspace{-4mm}
\begin{center}
\epsfxsize=3.0in
\leavevmode
\epsffile{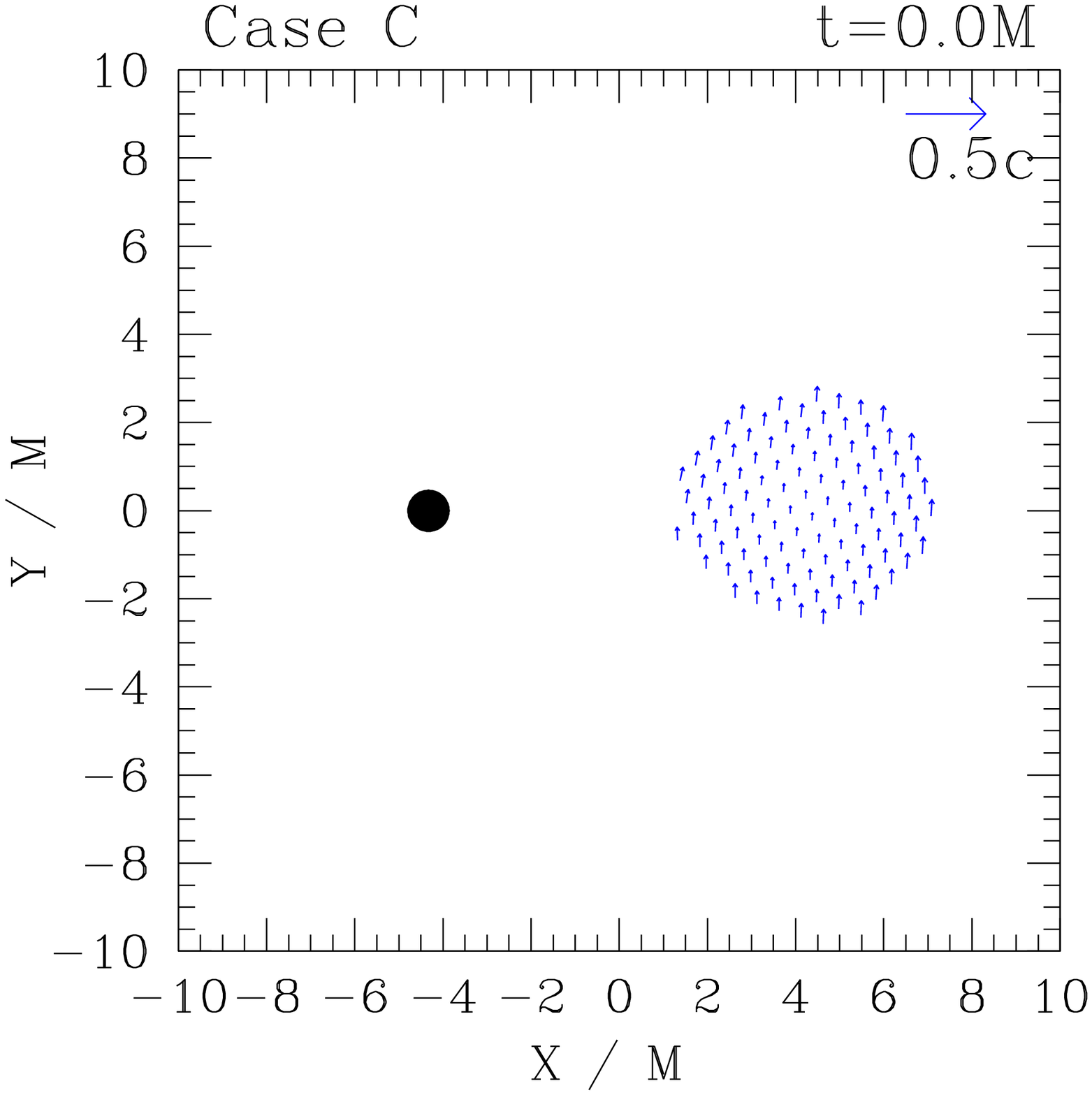}
\epsfxsize=3.0in
\leavevmode
\epsffile{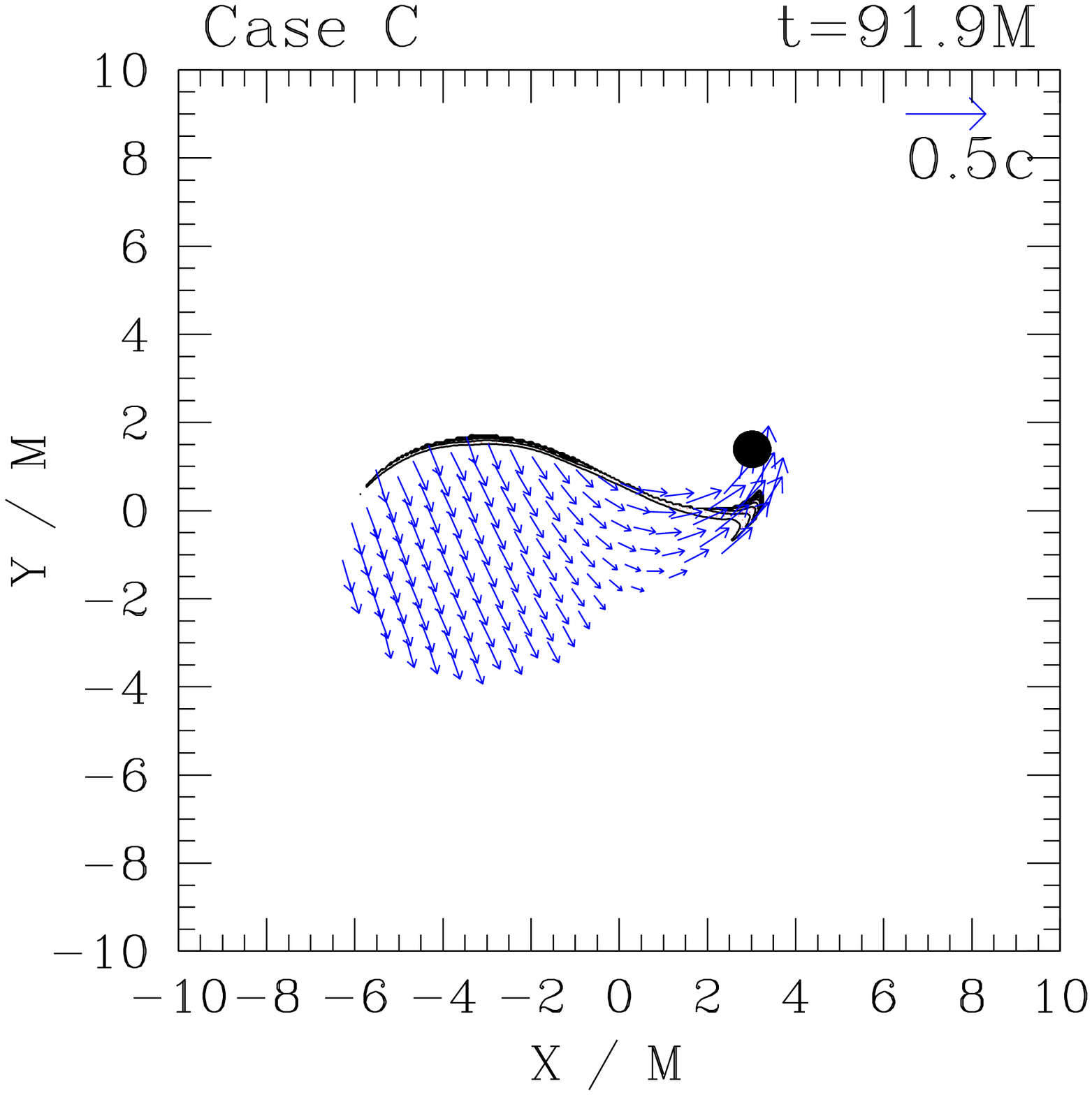}\\
\epsfxsize=3.0in
\leavevmode
\epsffile{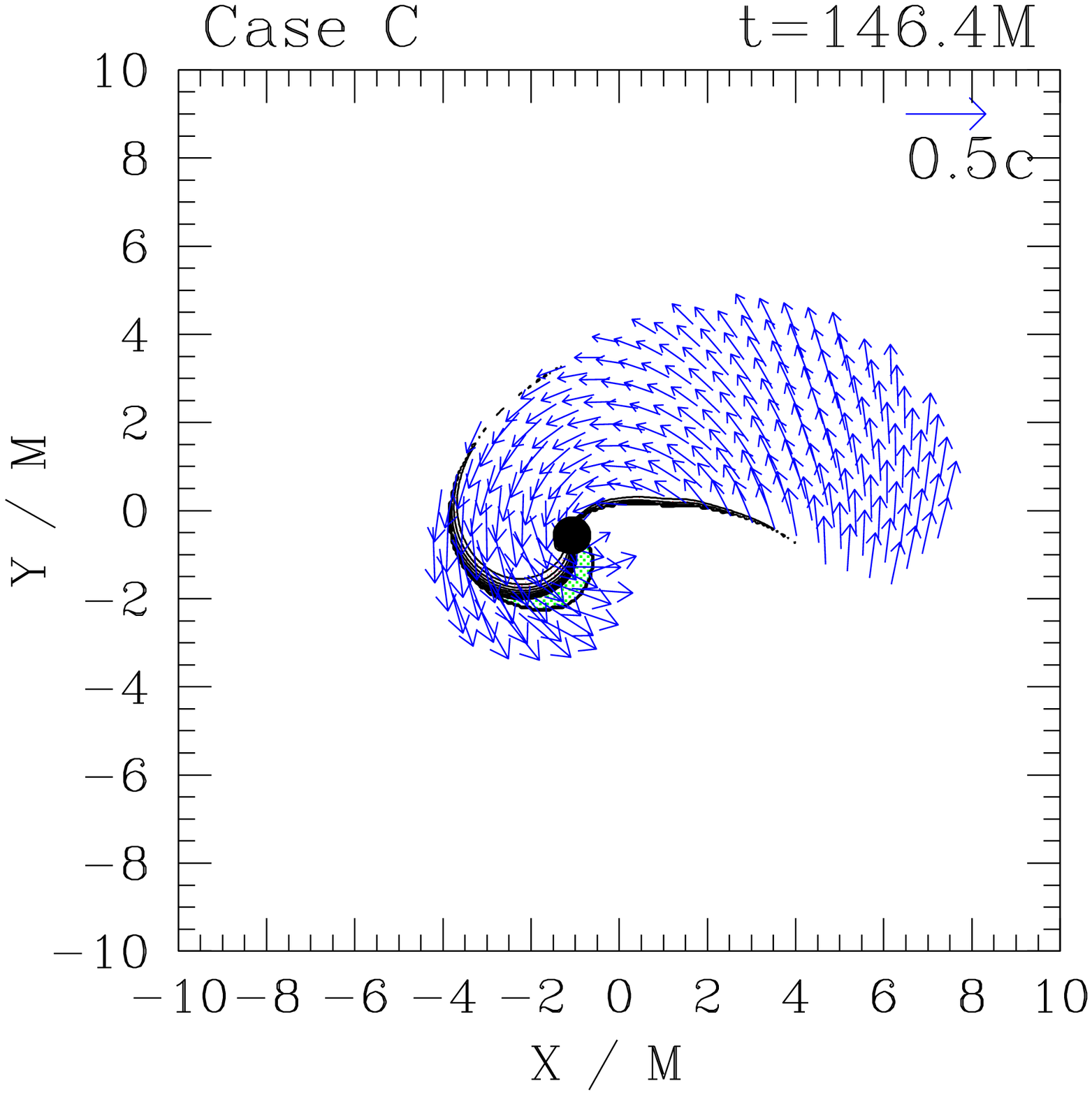}
\epsfxsize=3.0in
\leavevmode
\epsffile{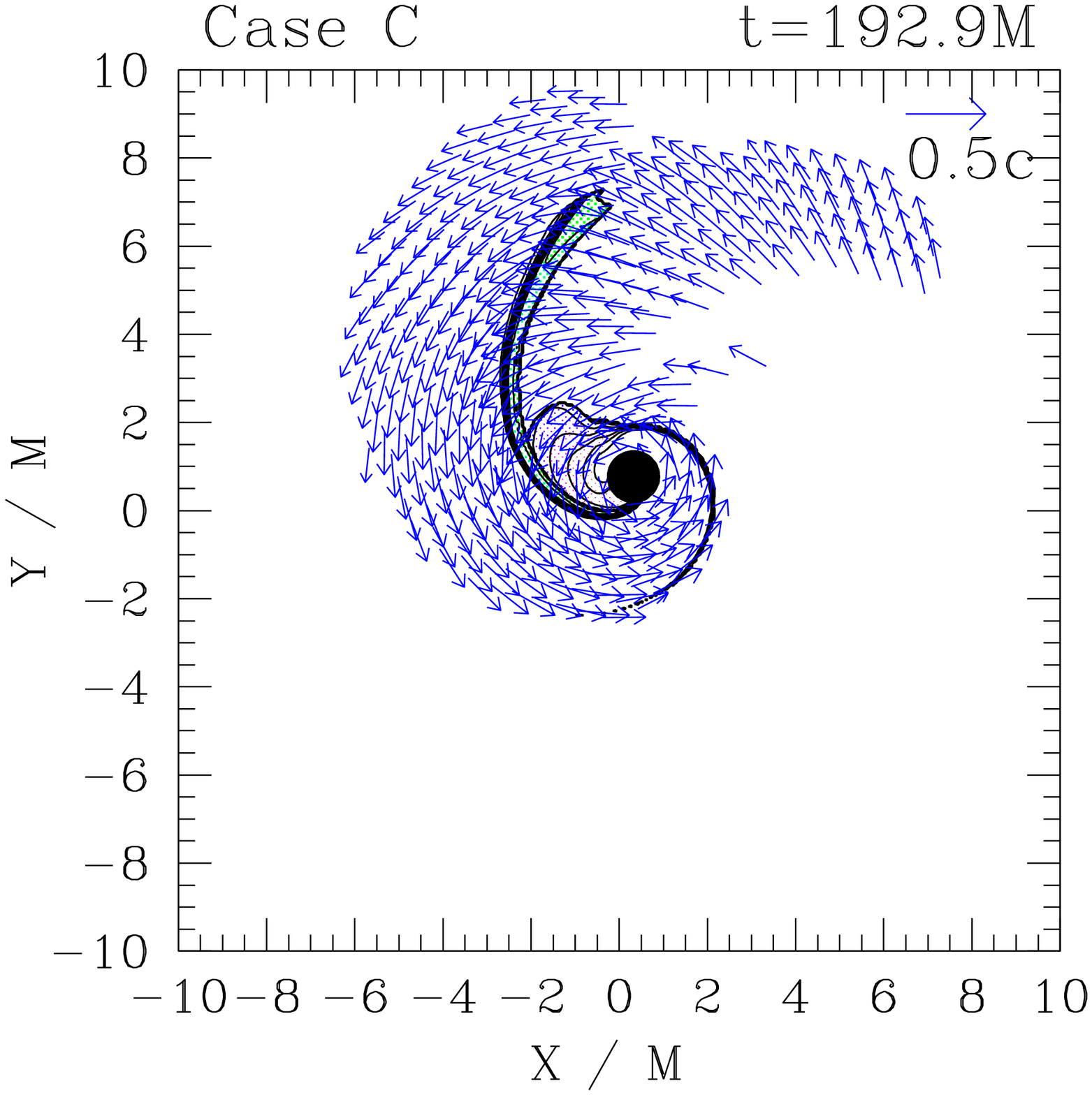}\\ %149
\epsfxsize=3.0in
\leavevmode
\epsffile{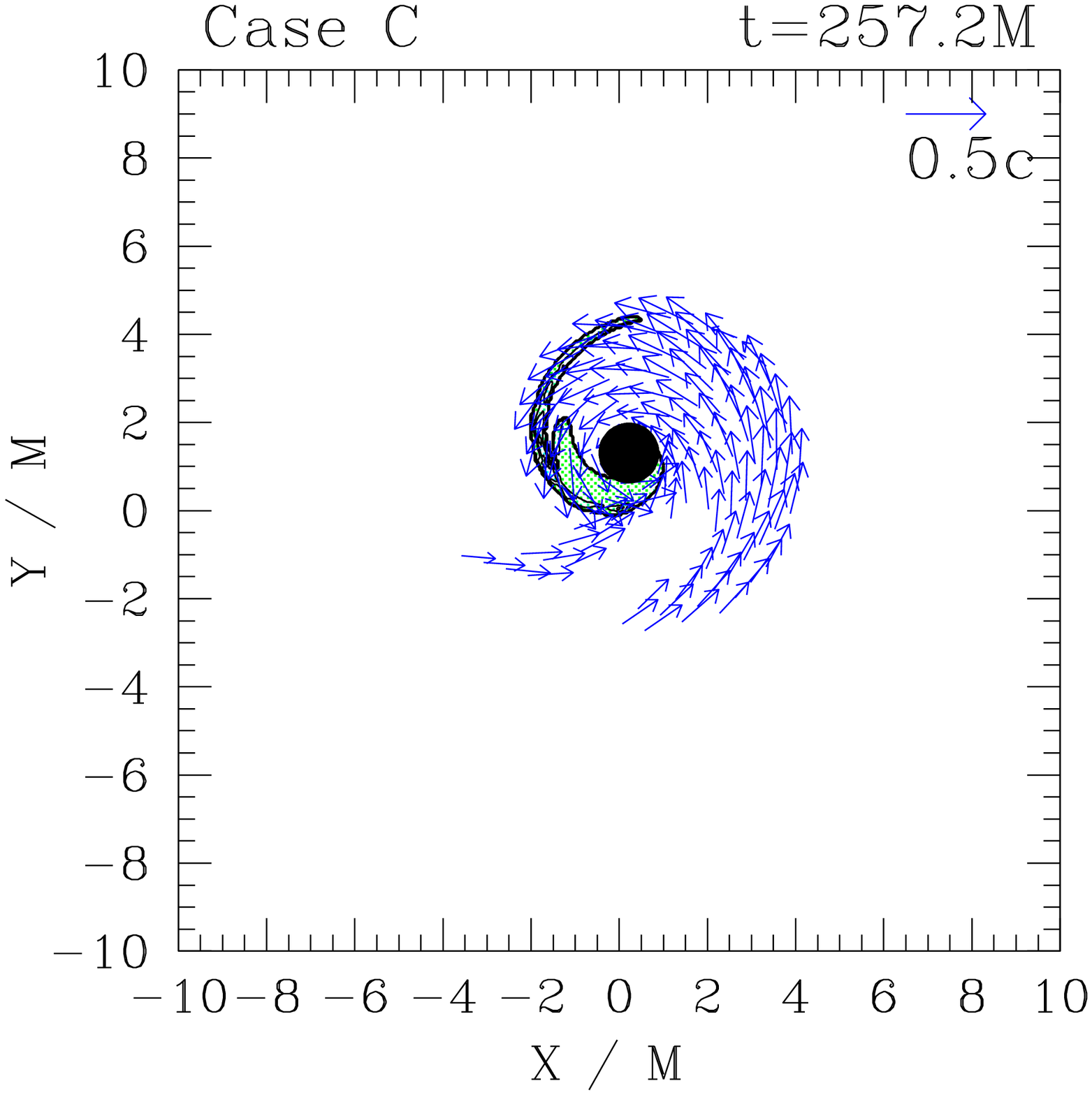} %245
\epsfxsize=3.0in
\leavevmode
\epsffile{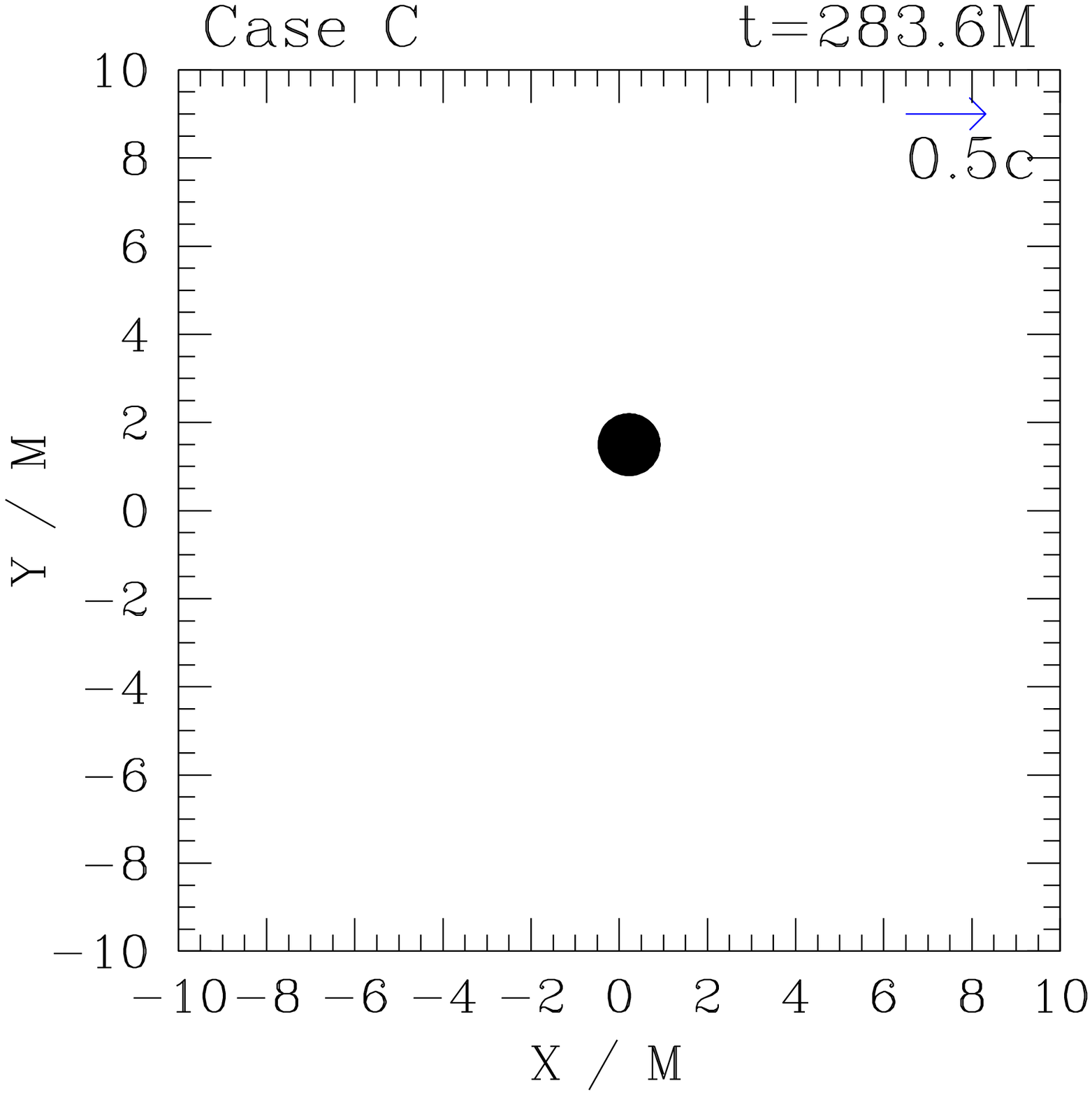} %327
\caption{Snapshots of equatorial contours of the polytropic constant
   $K$ at selected times for case C.  Contours are spaced linearly,
   so that $\Delta K=0.89$ from $K=2$ to $K=10$.  Values of $K>1$
   result from shock heating; $K=1$ for adiabatic flow.}
\label{fig:temp_c2}
\end{center}
\end{figure*}

We model our initial NS by a polytropic EOS $P=\kappa \rho_0^{\Gamma}$ 
with $\Gamma=2$. During the evolution, shocks develop and the matter 
heats up, resulting in an increase in the parameter
$K \equiv P/(\kappa \rho_0^{\Gamma})$ from its initial value of unity.
Fig.~\ref{fig:temp_c2} plots contours of $K$ at
different points in time for case C.  Notice that $K$ is clearly
larger than unity in the low-density overshoot region (upper-left panel of
Fig.~\ref{fig:temp_c2}), and where the overshoot region smashes into
the higher-density funnel region (upper-right panel of
Fig.~\ref{fig:temp_c2}).  We also observe shock heating as the final
bit of NS matter is accreted (lower-left panel of
Fig.~\ref{fig:temp_c2}).

A variety of physical effects are not modeled in the post-shocked,
semi-degenerate, nonequilibrium nuclear matter arising in these
simulations.  For example, transport due to photon and neutrino
radiation is not modeled, so accurate measurements of temperature $T$
are not possible.  Moreover, we are not employing a realistic hot
nuclear EOS.  However, we can very roughly estimate $T$ from the
specific energy density $\epsilon$.  For a polytropic equation
of state, the ``cold'' contribution $\epsilon_{\rm cold}$ is
\begin{equation}
\epsilon_{\rm cold} = -\int P_{\rm cold}
d(1/\rho_0) = \frac{\kappa}{\Gamma - 1} \rho_0^{\Gamma-1},
\end{equation}
where $P_{\rm cold} \equiv \kappa \rho_0^{\Gamma}$.  We now
define the thermal contribution to the specific energy density as
$\epsilon_{\rm th}=\epsilon-\epsilon_{\rm cold}$ and compute the
thermal contribution according to
\begin{eqnarray}
\epsilon_{\rm th} & = & \epsilon - \epsilon_{\rm cold} =
\frac{1}{\Gamma - 1} \frac{P}{\rho_0} - \frac{\kappa}{\Gamma - 1}
\rho_0^{\Gamma-1} \nonumber \\
& = & (K - 1) \epsilon_{\rm cold},
\end{eqnarray}
where we have used (\ref{eq:eos}) to express $\epsilon$ in terms of $P$ and 
$\rho_0$.  

To estimate $T$, we assume that we can model the  temperature dependence of
$\epsilon_{\rm th}$ as 
\begin{equation} 
\epsilon_{\rm th} = \frac{3kT}{2m_n} +
f\frac{aT^4}{\rho_0}
\end{equation}
(compare \cite{PWF}), where $m_n$ is the mass of a nucleon, $k$ is the
Boltzmann constant and $a$ is the radiation constant.  The first term
represents the approximate thermal energy of the nucleons, and the
second term accounts for the thermal energy due to radiation.  The
factor $f$ reflects the number of species of ultrarelativistic
particles that contribute to thermal radiation. When $T \ll 2m_e/k
\sim 10^{10}$K, where $m_e$ is the mass of electron, thermal radiation
is dominated by photons and $f=1$. When $T \gg 2m_e/k$, electrons and
positrons become ultrarelativistic and also contribute to radiation,
and $f=1+2\times (7/8) = 11/4$.  At sufficiently high temperatures and
densities ($T \gtrsim 10^{11}$K, $\rho_0 \gtrsim 10^{12}~{\rm g}~{\rm
cm}^{-3}$), neutrinos are generated copiously and become trapped, so,
taking into account three flavors of neutrinos and anti-neutrinos,
$f=11/4 + 6\times (7/8) = 8$. For a heated region in the tidally
disrupted NS, which possesses a 
typical density of $5\times10^{13}$g~cm$^{-3}$, and $K\sim7$, we
obtain a temperature of $T\sim2\times10^{11}$K.  Although
these temperatures are sufficient to produce copious neutrino
emission, our low disk mass would limit the overall neutrino energy to
$E_{\nu}\lesssim 10^{49}$ ergs (following the approximate scalings
derived by \cite{SRJ} from numerical models of BH disks).  This limits
the total $\gamma$-ray annihilation energy to $\lesssim 10^{48}$ergs
assuming $10\%$ efficiency, which may not be sufficient to power a
SGRB.  We note, however, that SGRB production may not require a
long-lived massive disk, since the actual emission mechanism remains
poorly understood.  Instead, we merely require that sufficient thermal
energy be produced to power the burst itself.  So until a more
detailed model of SGRB generation from BHNS mergers is developed,
any assessment of these simulations regarding SGRBs is tentative at
best.

\begin{figure*}
\vspace{-4mm}
\begin{center}
\epsfxsize=3.5in
\leavevmode
\epsffile{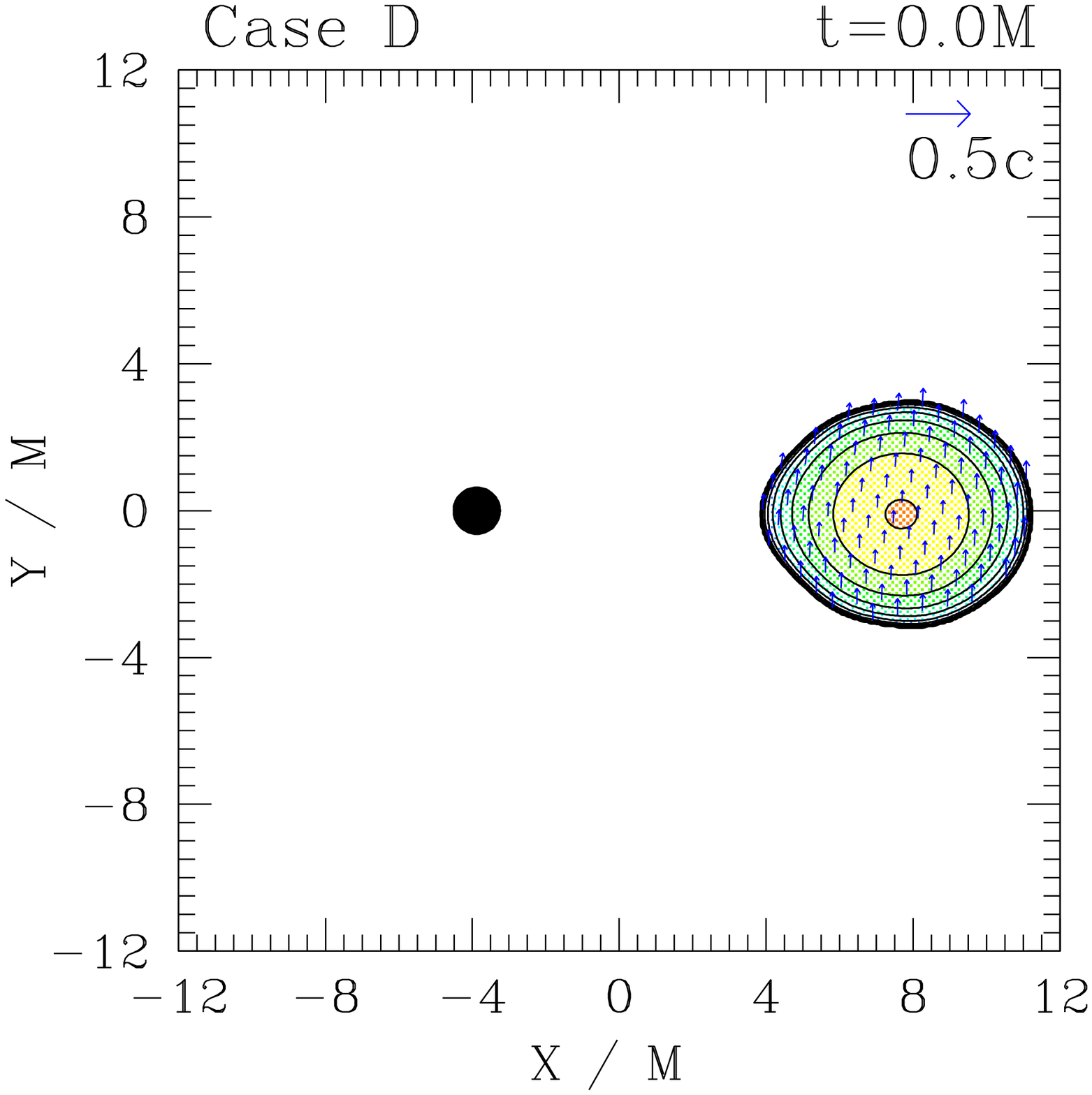}
\epsfxsize=3.5in
\leavevmode
\epsffile{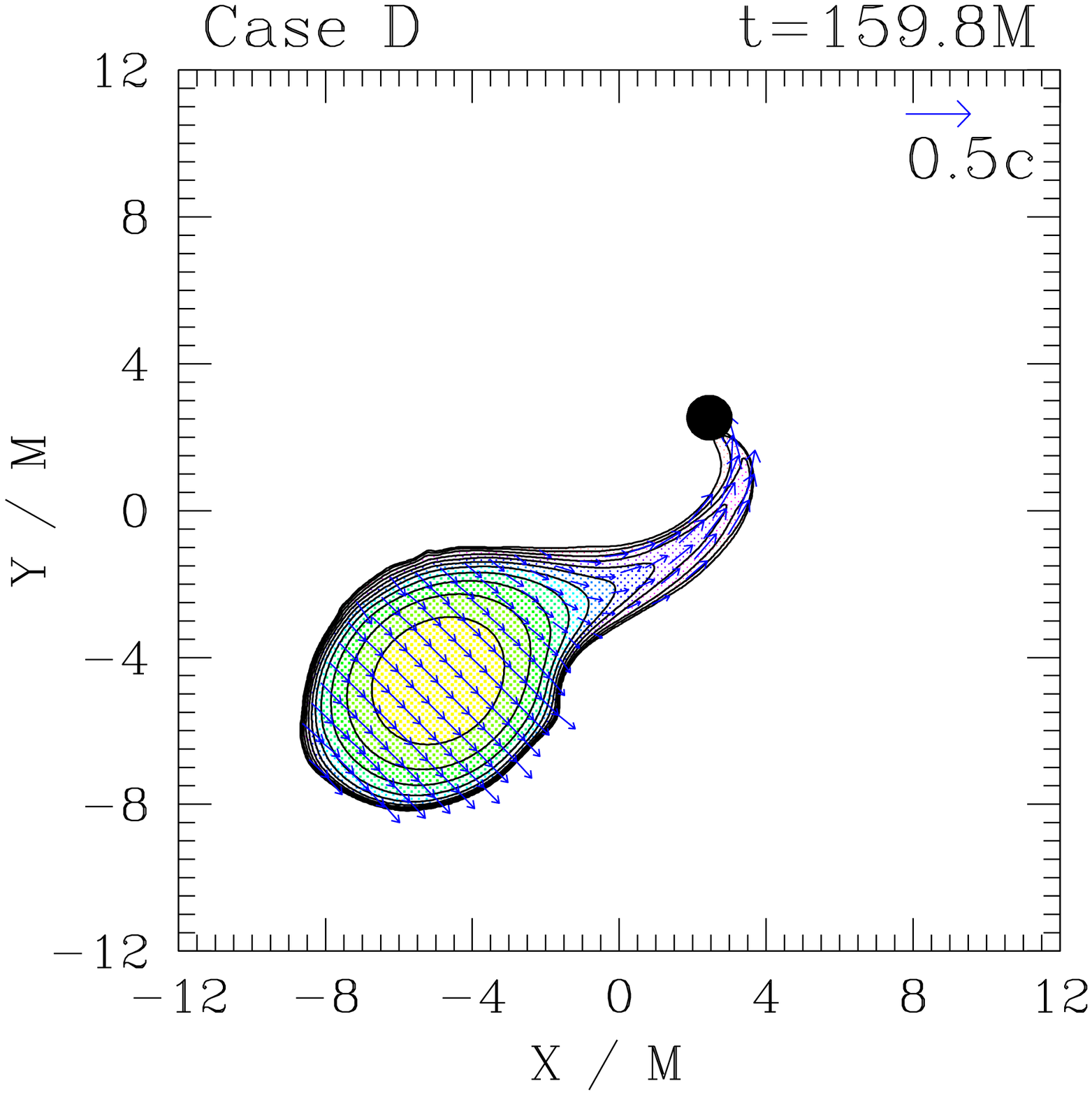}\\ 
\epsfxsize=3.5in
\leavevmode
\epsffile{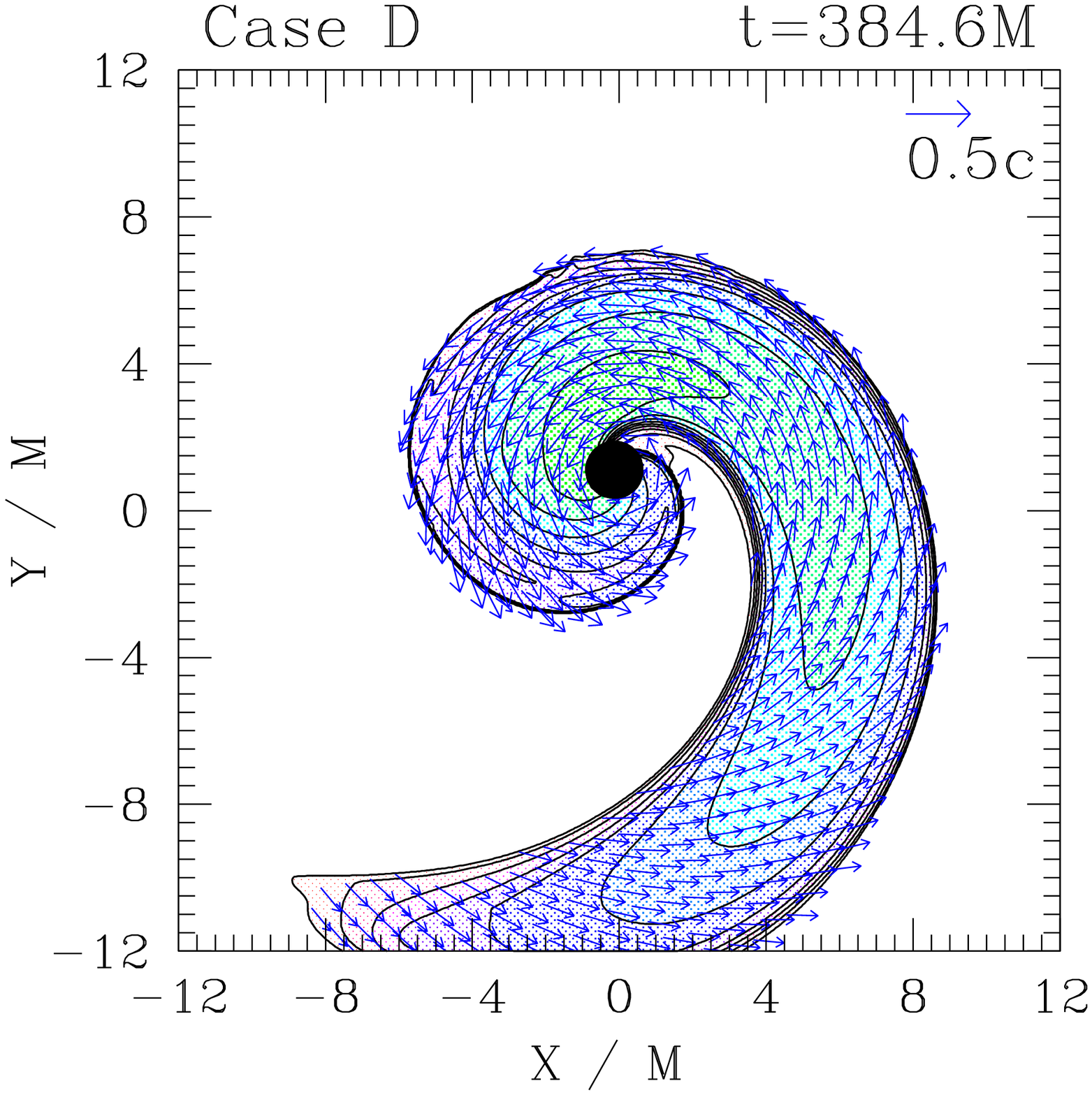} 
\epsfxsize=3.5in
\leavevmode
\epsffile{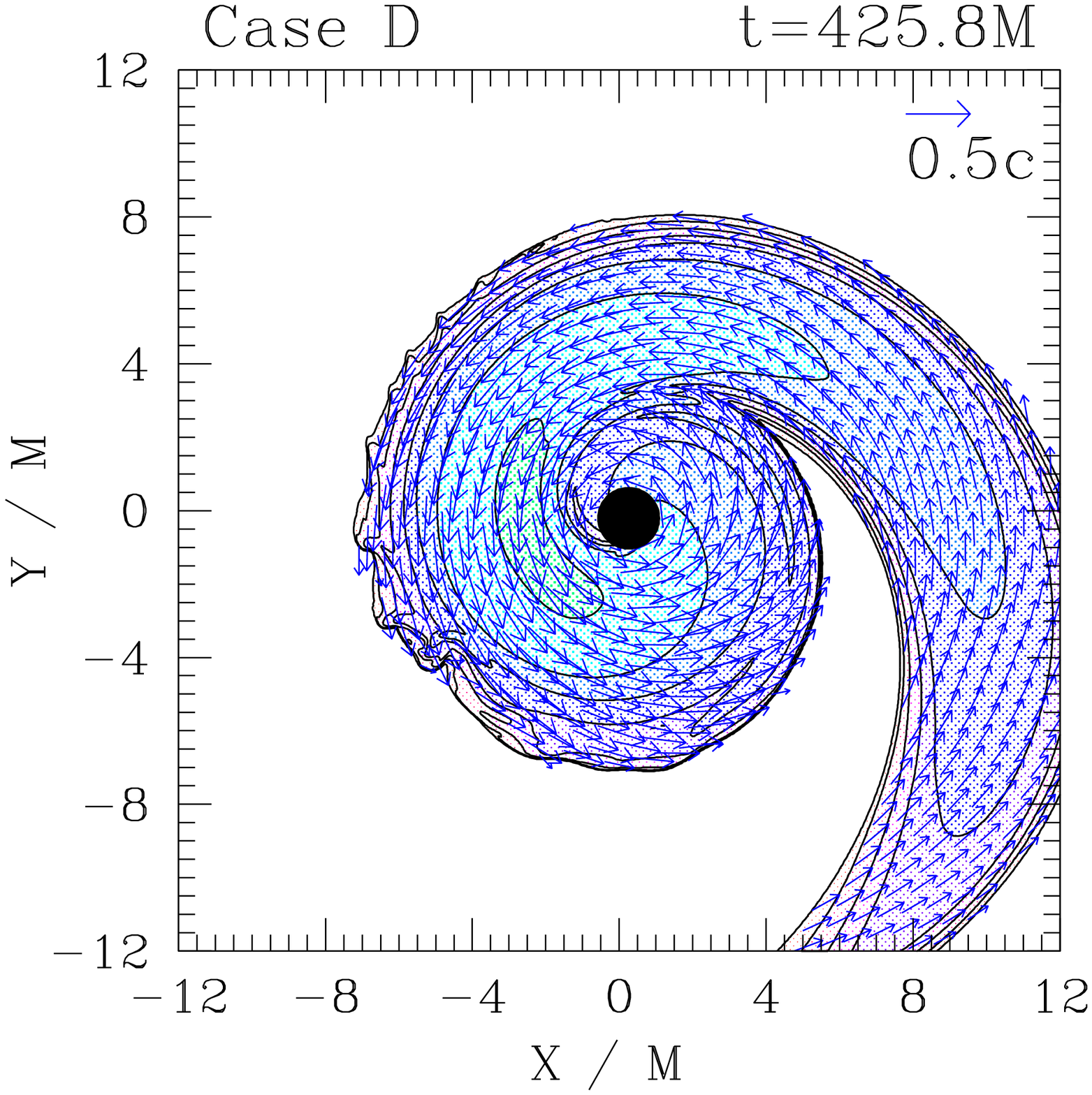} 
\caption{Snapshots at selected times from case D, the low-compaction,
   $q=2$ mass ratio case.  The contours represent
  the density in the orbital plane, plotted logarithmically with four
  contours per decade, with greyscaling added for clarity.  Arrows
  represent the velocity field in the orbital plane.  The minimum
  contour value in each frame is $\kappa \rho_0$(min)$=10^{-4}$,
  or $\rho_0$(min)$=1.6\times 10^{12} (1.4M_\odot/M_0)^2$g
  cm$^{-3}$.  The maximum initial NS density is $\kappa \rho_0 = 0.058$.
  We specify the black hole AH interior in each snapshot with a
  filled black circle.  In cgs units, the total ADM mass for this case is
  $M=2\times 10^{-5}(M_0/1.4M_\odot)$
  s$=6(M_0/1.4M_\odot)$km.}
\label{fig:xy_caseD1}
\end{center}
\end{figure*}

In Fig.~\ref{fig:xy_caseD1}, we plot snapshots of case D, the
low-compaction, $q=2$ mass ratio case.  At the time of first contact,
the accretion funnel is much narrower than in the high-compaction
cases of Fig.~\ref{fig:xy_seqABC}.  Further, unlike any of the
high-compaction cases, $\chi^{\text{funnel}}_{\angle}$ surpasses
$180^{\circ}$ at $t'\approx165M$ -- about $60M$ before the
highest-density region of the NS has been accreted.  After
$\chi^{\text{funnel}}_{\angle}=180^{\circ}$, a case C-like overshoot
develops (lower-left panel of Fig.~\ref{fig:xy_seqABC}), but instead
of smashing into the higher-density funnel and quickly falling in to
the BH, the overshoot gently merges with the funnel to create a
short-lived disk-like structure (lower-right panel of
Fig.~\ref{fig:xy_seqABC}) that later falls into the BH.

The final disk masses we measure for each run are listed in
Table~\ref{table:results}, along with an estimate of the final (Kerr)
BH spin.  To calculate the latter quantity, we take the ratio of the
polar to the equatorial circumference of the apparent horizon,
$C_r\equiv C_p/C_e$ and use Eq.~(5.3) of~\cite{bhspin} to solve for
the dimensionless spin $\tilde{a}\equiv a/M_{\mathcal{H}}$, where
$M_{\mathcal{H}}=(M_{\rm
  irr}/\tilde{a})\sqrt{2(1-\sqrt{1-\tilde{a}^2})}$ is the Kerr BH ADM
mass:
\begin{equation}
C_r=\frac{1+\sqrt{1-\tilde{a}^2}}{\pi}E\left(-\frac{\tilde{a}^2}{(1+\sqrt{1-\tilde{a}^2})^2}\right).\label{eq:spinbh}
\end{equation}
Here $E(x)$ is the complete elliptic integral of the second kind.  In
all cases, we find disk masses of less than $2.8\%$ of the original NS
mass.  The final (Kerr) BH spin is roughly $a/M=0.5$, $0.64$, $0.8$,
and $0.5$ for cases A, B, C, and D, respectively.  Here we use our
finding that $M=M_{\mathcal{H}}$ to good approximation.  The 
first of these agrees well with similar results of ST; this is the
only case for which a meaningful comparison is possible, given the
adopted mass ratios.

\begin{table*}
\caption{Final results from each of our simulations.  We list the
  fractional rest (baryon) mass outside the horizon $f_{\rm out}\equiv
  M_0(r>r_{\rm AH})/M_0$ and the dimensionless spin of the BH
  $a/M$ [see Eq.~(\protect\ref{eq:spinbh})] at the end of our
  simulation.  Also shown are the radiated energy, angular
  momentum, and linear velocity ``kick'' resulting from GW emission,
  the former two normalized to the binary's initial total ADM mass and
  the latter in km/s.  For the GW quantities, entries without (with)
  parentheses are derived from the Z-M ($\psi_4$) formalism. Note that in case~D, the GW data are not accurate enough to obtain 
reliable estimate of $\Delta E_{\rm GW}$, $\Delta J_{\rm GW}$ and the 
kick velocity.}
\begin{tabular}{|c c c c c c|}
  \hline
  Case & $f_{\rm out}$  &
  $a/M$ & $\Delta E_{\rm GW}/M$ & $\Delta J_{\rm GW}/M^2$ &
  Kick velocity (km/s) \\
  \hline
  A1-lo & $<$1\% & $\approx  0.52$ & 0.60\% (0.70\%) & 5.2\% (6.3\%) & 21 (21) \\
  A1-med & $<$1\% & $\approx 0.52$ &  0.79\% (0.75\%) & 6.6\% (6.2\%) & 39 (20)  \\
  A1-hi & $<$1\% &$\approx  0.52$ & 0.65\% (0.77\%) & 5.4\% (6.4\%) & 46 (19) \\
  A1-farbc & $<$1\% &$\approx  0.52$ & 0.74\% (0.74\%) & 6.0\% (6.1\%) & 15 (16)  \\
  A2 &$<$2\% & $\approx 0.52$ & 0.72\% (0.66\%) & 7.3\% (6.5\%) & 49 (24) \\
  A3 & $<$2.8\%  &$\approx 0.48$ & 0.67\% (0.86\%) & 7.9\% (5.1\%) & 18 (30)  \\
  B & $<$1\% & $\approx  0.64$ & 0.59\% (0.53\%) & 6.2\% (6.2\%) & 67 (48)  \\
   C  & $<$1\% & $\approx 0.80$ & 0.39\% (0.32\%) & 5.4\% (4.7\%) & 22 (25) \\
  D & $<$1\% & $\approx  0.5$  & - & - & - \\
  \hline
\end{tabular}
\label{table:results}
\end{table*}

\subsection{Gravitational Wave emission} \label{sec:gw_emission}

The gravitational wavetrain from a compact binary system may be
separated into three qualitatively different parts: the inspiral,
merger, and ringdown.  We describe each briefly before discussing our
numerical results.

During the inspiral phase, which takes up most of the
binary's lifetime, GW emission circularizes the orbit and gradually
reduces the binary separation.  At the large binary separations during
the inspiral stage, finite-size effects associated with the NS are
unimportant, and post-Newtonian (PN) techniques are sufficient to describe the
evolution.  At present, the binary orbit dynamics 
is determined to 3.5PN order (e.g.~\cite{BFIJ}) and the corresponding 
GW emission is computed to 2.5PN order~\cite{abiq04,kbi07} 
(but see also~\cite{kidder07}). Even after
finite-size corrections become relevant, quasi-equilibrium sequences
allow for a determination of the binding energy as a function of
orbital frequency, from which the GW energy spectrum $dE/df$ may be
calculated, following the techniques described in \cite{FGRT}.  This
method was used in \cite{TBFS07a} to determine the approximate energy
spectrum from the sequences we use as initial data in this work.

Once the binary nears the ISCO, or the point where tidal disruption
begins, the orbit decays rapidly and the GW emission changes
character.  In particular, deviations from point mass behavior
typically result in a sharp decline in the energy spectrum. We note
that these ``break frequencies'' marking the onset of instability
systematically occur at lower frequencies for BHNS than for BHBH
binaries (see discussion in~\cite{Ajith}), especially in cases where
tidal disruption occurs.  This can be seen clearly in \cite{TBFS07b},
noting that the tidal disruption branches do not exist for BHBH
systems.

Finally, we expect a phase of quasinormal ringing of the BH, since it
is distorted by the merger.  This emission typically results in a
higher frequency peak in the energy spectrum, with an amplitude
determined by the total distortion induced on the BH by the merger.

\begin{figure}
\epsfxsize=3.4in
\leavevmode
\epsffile{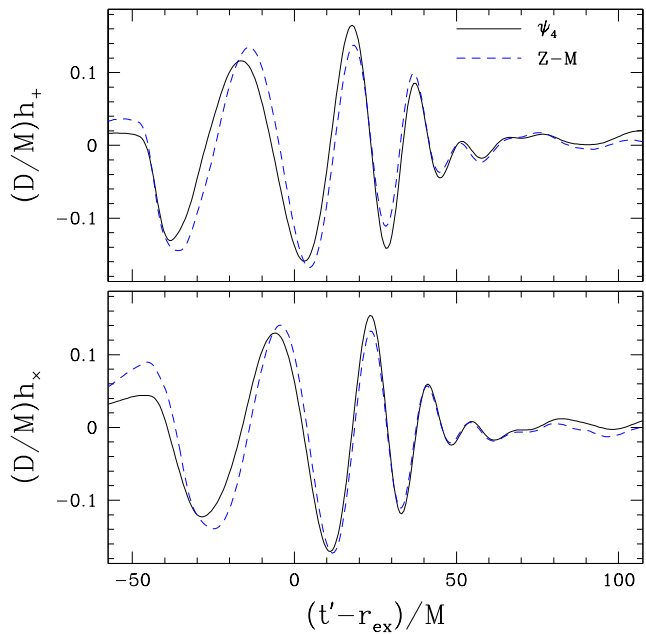}
\caption{Gravitational wave signal from case A1-hi, calculated using
  $\psi_4$ (solid line) and Z-M (dashed line). We show both polarizations as seen by
  an observer looking down the initial polar axis, $h_+$ (top panel)
  and $h_\times$ (bottom panel).  The scale factor $D$ is the distance
  to the binary.}
\label{fig:waves_runa1hi}
\end{figure}

In Fig.~\ref{fig:waves_runa1hi}, we plot the GW strains along the
polar axis of the binary for case A1-hi, using both the $\psi_4$
formalism (solid curves; Eq.~(\ref{eq:psi4})) and the Z-M formualtion
(dashed curves; Eq.~(\ref{eq:moncrief})).  In both cases the waveforms
are extracted on a sphere of physical radius $r_{ex}=34.4M$, and modes
up to and including $l=4$ are used in the calculation.  
We add suitable integration constants when computing the waveforms from 
both Z-M (odd-parity modes) and $\psi_4$ formalisms 
to minimize offsets in the time-averaged $h_+$ and
$h_\times$.  We see that there is some disagreement at early times as
spurious gravitational radiation present in the initial data propagate
outward.  Once this ``junk'' radiation has left the numerical grid,
the two independent methods yield results that are in very good
agreement, even though they are calculated using different sets of
metric components. 

During the inspiral phase, the GW frequency and amplitude sweep upward
until the point at which the NS begins to be disrupted by the BH, at
$t'=0M$.  As accretion progresses , there is a gradual but steady
downturn in the amplitude while the GW frequency continues to sweep
upward.  Finally, from $t'=50M$ onward, after the vast majority of the
NS matter has been accreted, we see a very weak ringdown signal, at
amplitudes significantly less than those seen in either BHBH
(\cite{RIT1,God1,Jena_BBH}) or NSNS (\cite{ST}) mergers.

As discussed in Sec.~\ref{sec:caseA}, the numerical resolution has
little effect on the mass accretion rate in sequence~A.
Correspondingly, we find that the tidal disruption signature in
sequence~A waveforms is largely resolution-independent.  For
runs A1-hi, A1-med, and A1-low, we estimate the GW frequencies at
$t'-r_{ex}=0$ to be $M\Omega_{\rm GW}\equiv 2\pi Mf_{\rm GW}=0.188$,
$0.180$ and $0.190$, respectively.  These frequencies are slightly
higher than twice the orbital frequency value at tidal disruption
found in \cite{TBFS07a}, $M\Omega_{\rm orb}\approx 0.07$, as is expected
since first contact occurs slightly after the onset of tidal
disruption.  

\begin{figure}
\epsfxsize=3.4in
\leavevmode
\epsffile{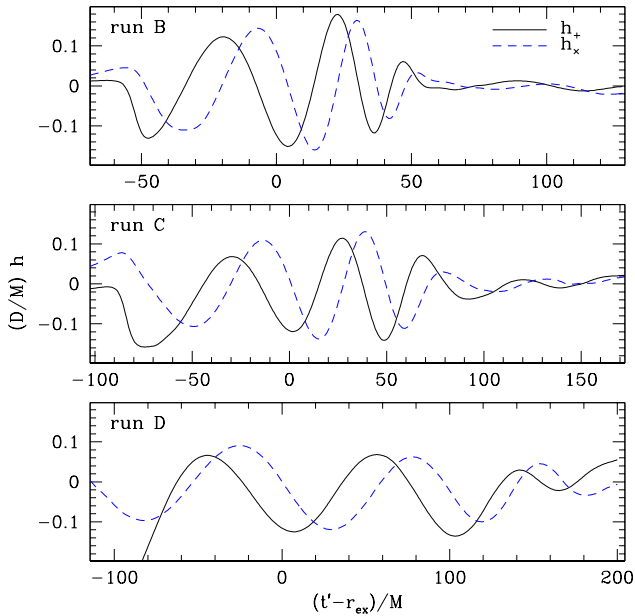}
\caption{Gravitational wave signal from cases B, C, and D, calculated using
  the Z-M formalism.  We show both polarizations, $h_+$ (solid)
  and $h_\times$ (dashed) along the polar axis of the binary.  The
  scale factor $D$ is the distance to the binary.}
\label{fig:waves_bcd}
\end{figure}

A similar pattern is observed in cases B and C (see
Fig.~\ref{fig:waves_bcd}).  The ringdown amplitude grows relative to
the overall signal strength as $q$ is reduced from $3$ (sequence~A) to
$1$ (case C).  However, in all cases the amplitude is significantly
smaller than the comparable BHBH ringdown signal; we discuss this
issue in more detail below.

The low compaction NS in case D implies a larger NS radius, so a
larger initial binary separation was required than for the other
cases.  Thus, case D required many more grid light-crossing times
until merger and ringdown.  As a result, late-time normalized
Hamiltonian constraint violation $||\mathcal{H}||$ increased to
$\approx8\%$, leading to an inaccurate late-time waveform.  We
therefore truncate the waveform after $t'-r_{ex}=200M$.  Based on our
analysis of sequence~A and the known high computational cost of
case~D, accurate simulations of case~D would require higher resolution
and more distant outer boundaries than is practical, given our
computational resources.  Though we can still evolve the matter and
trajectory of the BH reliably, we find that since the waveforms are
manifested as small perturbations on the background spacetime, they
are greatly affected by constraint violations.

We tabulate the GW energy loss $\Delta E_{\rm GW}$ and angular
momentum loss $\Delta J_{\rm GW}$, as well as the measured
kick velocity imparted to the BH in the rightmost columns of
Table~\ref{table:results}.  Quantities without parentheses are
derived from the Z-M formalism waveforms, and those in parentheses
are derived from our $\psi_4$-based waveforms. We compute $\Delta
E_{\rm GW}$, $\Delta J_{\rm GW}$, and the kick velocity at 5
radii in the range $\approx 31-37M$ for all cases except A1-farbc
(where the radii span $\approx 47-76M$). The values shown in
Table~\ref{table:results} are obtained by Richardson extrapolation of
the data to $r\rightarrow \infty$.  In general, we find good agreement
between the two GW measurement methods, especially for the case with
more distant outer boundaries, which satisfied the constraints best at
late times. Based on the variations of results with different
resolutions (for cases A1), different GW extraction radii and in the
two GW extraction methods, we estimate that our tabulated $\Delta
E_{\rm GW}$ and $\Delta J$ is accurate to about 20\%, whereas the
error in kick velocity may be as much as 50\%.  In case~D, the GW data are
not accurate enough to provide reliable data for energy, angular
momentum losses and kick velocity.

Compared to previous simulations of merging BHBH systems with
the same mass ratio \cite{JenaQ}, we find that while the radiated
energies, angular momenta and kick velocities are significantly lower in
our runs because tidal disruption suppresses the GW signal, the final
BH spins are comparable, within our uncertainties, to BHBH values.

\begin{figure}
\epsfxsize=3.4in
\leavevmode
\epsffile{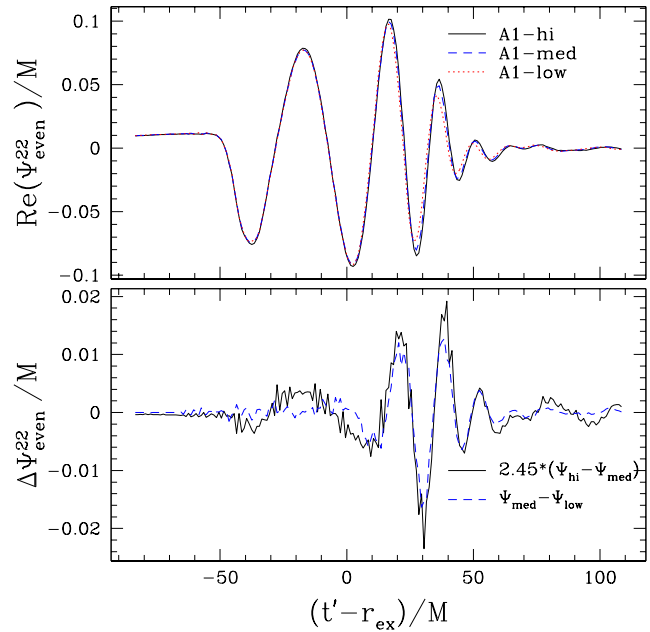}
\caption{Numerical convergence of gravitational wave signals for cases
   A1-hi, A1-med, and A1-lo.  In the top panel, we show the $l=m=2$
   component of the even-parity mode $\Psi^{22}_{even}$ for the three
   waveforms, noting that while they remain in phase with each other,
   we see overall amplitude differences on the order of several
   percent.  In the bottom panel, pairwise differences between
   the waveforms are plotted, with the higher-resolution pair rescaled
   to demonstrate second-order convergence.}
\label{fig:waveconv_runa1}
\end{figure}

In a previous work \cite{EFLSB}, we showed that our $\psi_4$
measurements converged to second-order with numerical resolution.  In
Fig.~\ref{fig:waveconv_runa1}, we perform a similar demonstration, but
with the Z-M formalism.  In this figure, we plot the real component
of the $\Psi^{22}_{even}$ mode for cases A1-hi, A1-med, and A1-low,
which differ only in the numerical grid spacing.  The three waveforms
are plotted in the top panel and show good agreement.  Notice that the
waveform amplitudes are weakly dependent on the numerical resolution,
but only at the level of a few percent for our higher resolution runs.
In the bottom panel, we show differences between pairs of runs,
rescaling the higher-resolution case by a numerical factor that
assumes second-order convergence,
$(166^{-2}-220^{-2})/(220^{-2}-270^{-2})=2.25$, finding agreement.
Although our spatial differencing scheme for
the fields is fourth-order accurate, our HRSC scheme in unshocked
regions is only second-order accurate.  This, along with the
appearance of shocks (which are only first-order convergent) limits
the convergence order of our waveforms over time.  As a result, while
we can perform fourth-order time integrations with our code, we
generally prefer second-order time integration since it is faster and
does not result in a significant loss of accuracy.

\begin{figure}
\epsfxsize=3.4in
\leavevmode
\epsffile{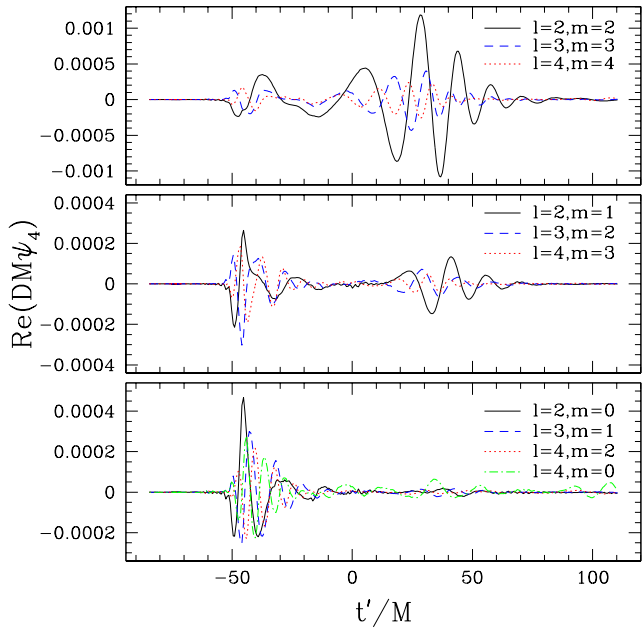}
\caption{Mode decompostition of the real component of $\psi_4$ for
  case A1-hi.  In the top panel, we show the dominant $l=m$ modes.  In
  the middle panel, we plot the modes that satisfy $l=m+1$, and in the
  bottom panel we show several other modes.}
\label{fig:wavemodes}
\end{figure}

Although the $l=2$, $m=2$ mode is the dominant component of the
radiation, we measure all spin-weighted spherical harmonic components
up to and including $l=4$.  In Fig.~\ref{fig:wavemodes}, we show the
mode decomposition of $\psi_4$ as a function of time for all
non-negligible contributors.  In the top panel of the figure,
components satisfying $l=m$ are plotted, including the dominant
$l=m=2$ mode.  Notice that the modes satisfying $l=m+1$ in the middle
panel possess an amplitude that is at most $15\%$ of the total strain
at any given moment.  For completeness, we plot some of the other
significant modes in the bottom panel, noting that while they are
present in the initial passage of ``junk'' radiation, they play little
or no role at later times.

To study the detectability and qualitative features of our computed
gravitational wavetrains, we calculate the effective GW wave strain in
frequency space using Eq.~(\ref{eq:heff}).  We find that if an FFT is
performed on these GW signals without modification, the initial burst
of spurious junk radiation contributes significantly to the signal,
and the finite initial amplitude introduces a strong aliasing signal
across the entire frequency domain of interest.  To fix these
problems, we perform ``surgery'' between our numerical GW signal, at a
point in time where the initial junk has passed through the GW
extraction surface, and a post-Newtonian signal with the same $q$.  We
generate the restricted PN waveform following
the same techniques as in, e.g., \cite{Ajith}.  Defining $v\equiv (\pi
M f_{\rm GW})^{1/3}$, we take as an initial condition the value of $v$
computed from the phase evolution of our numerical waveform and evolve
backwards in time the following set of equations,
\begin{eqnarray}
\frac{d\varphi}{dt}&=&\frac{2v^3}{M},\\
\frac{dv}{dt}&=&-\frac{1}{M}\frac{F(v)}{dE(v)/dv},
\end{eqnarray}
where the binding energy per unit mass $E(v)$ and
radiation flux $F(v)$ are taken from the PN calculations of
\cite{BFIJ}. We modify the amplitude of the PN signal to minimize
aliasing, but the relative correction is in all cases less than 2\%.
Mismatches in the amplitude, frequency, and frequency sweep rate
appear as oscillations in the energy spectrum near the surgery
frequency.  As a result, any peaks and troughs appearing at comparable
frequencies in the energy spectrum should be viewed with skepticism
unless demonstrated to be robust with respect to the surgery
procedure.  At both the beginning and end of the combined waveform, we
add exponential damping terms to reduce aliasing, but this operation
does not add power at frequencies of interest.

\begin{figure}
\epsfxsize=3.4in
\leavevmode
\epsffile{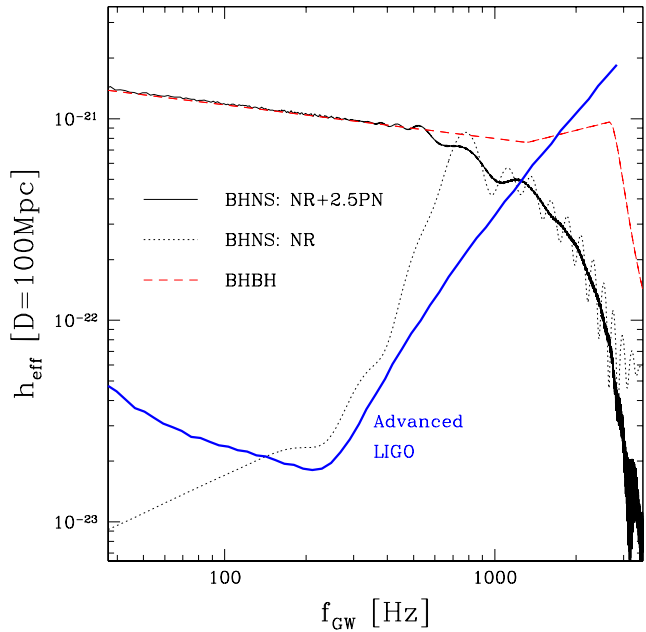}
\caption{Gravitational wave spectrum for the case A3 BHNS merger
  compared to a BHBH merger with the same masses.  The solid curve
  shows the combined waveform found by attaching the restricted PN
  waveform to our numerical signal, while the dotted
  curve shows the contribution from the latter only, demonstrating the
  expected aliasing behavior resulting from FFTs of discontinuous
  functions.  The dashed curve is the analytic fit derived by
  \protect\cite{Ajith} from analysis of multi-orbit BHBH inspirals,
  which maintain significantly more power at higher frequencies.  The
  heavy solid curve is the effective strain of the advanced LIGO
  detector, defined such that $h_{\rm LIGO}(f)\equiv \sqrt{fS_h(f)}$.
  To set physical units, we assume a NS rest mass of $M_0=1.4M_\odot$.}
\label{fig:heff_runa}
\end{figure}

The result of this operation for run A3 is shown in
Fig.~\ref{fig:heff_runa}, where units are set by assuming a NS 
rest mass of $1.4M_\odot$.  We plot the effective strain computed from
both our combined waveform and our numerical signal alone.  Notice
that the signal follows the point-mass power law behavior up to
frequencies of approximately $f_{\rm GW}=600-800~{\rm Hz}$, at which
point disruption of the NS and its subsequent accretion dominates the
signal.  At frequencies above $f_{\rm GW}=1~{\rm kHz}$, however, there
is extremely little power in the waveform since the 
ringdown signal is so weak.  These results are consistent with what we
expect from the quasi-equilibrium configurations of \cite{TBFS07b},
who find that $f_{\rm GW}=\Omega_{\rm orb}/\pi\approx 800~{\rm Hz}$ for the
configuration in question ($Mf_{\rm GW}=0.022$ in dimensionless
units).  They also agree roughly with results of \cite{ST07}, who find
a similar pattern of steep decline above the
tidal disruption value.

Each BHNS merger spectrum is compared to a BBH merger spectrum, taken
from Eqs.~(4.12)-(4.19) of \cite{Ajith}, noting that
$h_{\rm eff}(f)\propto fA(f)$ in their notation.  The comparison is
performed using a binary with the same masses as our BHNS case. Both
of these curves lie above the advanced LIGO sensitivity band $h_{\rm
  LIGO}(f)\equiv \sqrt{fS_h(f)}$, which we have
taken from \cite{ADVLIGO}.  This result assumes a distance to either
source of $D=100~{\rm Mpc}$, the distance required to reach one merger
per year assuming an overall rate of 10 mergers per megayear per Milky
Way-equivalent galaxy (and a density of these 
of $0.1~{\rm gal/Mpc}^3$) \cite{BBR}.  This distance is roughly that of the Coma
cluster, and approximately five times the distance to the Virgo cluster.
The difference in wave signal between BHBH and BHNS mergers is 
present in the advanced LIGO frequency band, but only marginally.  It
is clear that for more significant measurements of the difference
between BHNS and BHBH inspirals and mergers, it would be advantageous
to make use of narrow-band detection techniques with advanced detectors.

\begin{figure}
\epsfxsize=3.4in
\leavevmode
\epsffile{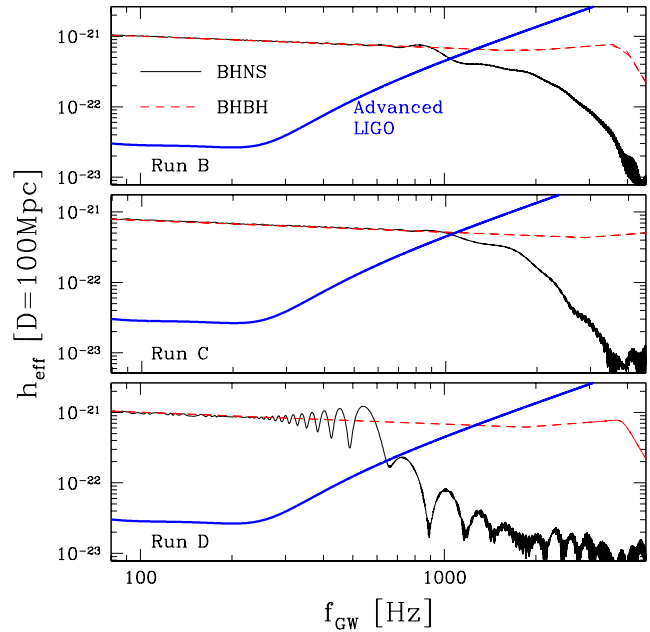}
\caption{Gravitational wave spectra for the cases B, C and D BHNS
  mergers compared to BHBH merger.  Conventions are as in
  Fig.~\ref{fig:heff_runa}.}
\label{fig:heff_bcd}
\end{figure}

Fig.~\ref{fig:heff_bcd} contains plots of the GW spectrum for cases B,
C, and D.  As the mass ratio and NS compaction is varied, we see the
expected differences in the apparent ``break frequency'' marking tidal
disruption.  This frequency may be estimated by $M f_{\text{break}}
\approx (M/d_{\text{tid}})^{3/2}/\pi \approx
{\cal C}^{3/2} (1+q)\sqrt{(1+q)/q}/\pi$, where Eq.~(\ref{eq:d_tid}) has been used for
$d_{\text{tid}}$.  This formula is consistent with Eq.~(25) of
\cite{TBFS07b}.  As the value of $q$ is lowered from 3.0 to 1.0, the
break frequency rises.  In addition, when we lower the compaction, we
see a large decrease in the break frequency.  These results, which
agree well with the empirical scalings derived in \cite{TBFS07b}, lend
credence to the idea that if the individual masses of the binary
components can be derived from the inspiral waveforms, the GW break
frequency should provide a relatively sensitive measurement of the NS
radius.  When combined with observations of the lower-frequency
spectrum, these inferences may provide additional constraints on the
NS structure, including limits placed on the tidal Love number $k_2$
(following the techniques described in \cite{FlanHind} for NSNS
mergers).

\section{Discussion}\label{sec:discussion}

In this paper we present our first fully self-consistent, dynamical simulations
of relativistic BHNS binaries.   
% Old, alternate version
%We have applied the techniques we developed and validated in
%\cite{EFLSB} to ``fill'' in excised BHNS CTS initial data we generated
%in \cite{TBFS07a}, focusing on irrotational BHNS cases with mass
%ratios in the range $q=1-3$.  We then apply our field+matter evolution
%techniques described in \cite{FBEST} to evolve these data through
%inspiral, merger, and ringdown.
We use results and evolution techniques that we
have previously developed in preparation for these
simulations, including the initial data of \cite{TBFS07a}, the
``filling'' of the black holes in these initial data \cite{EFLSB} and
the treatment of relativistic hydrodynamics in the context of the
moving puncture method \cite{FBEST}.  Here we focus on irrotational
BHNS binaries with mass ratios between $q = 1$ and 3. 

For the cases studied here, we find no more than $\sim3\%$ of the
original NS matter remaining outside the BH at the end of the
simulations.  Such small disk masses lend support to the semi-analytic
arguments presented in \cite{MCMiller}, which suggested that virtually
the entire NS will be accreted promptly by the BH.

The simulations of ST, on the other hand, suggest larger disk masses
than ours.  The reason for this discrepancy remains unclear, but we
suspect it may be caused by different initial data -- both choices of
parameters and/or computational approach.  We note that the mass
accretion process depends on the initial binary 
separation (compare runs A1, A2 and A3 in Fig.~\ref{fig:mass_runs_a}),
suggesting that the disk mass depends rather sensitively on the details
of the initial data.  However, we cannot rule out that this dependence
is instead a numerical artifact, caused by numerical errors that are
growing over time due to outer boundaries that are too close to the
strong-field region and other effects.  Since we model NSs as
$\Gamma=2$ polytropes, and the disk mass is likely to depend
sensitively on the NS EOS, firm conclusions about BHNS mergers as SGRB
progenitors remain uncertain.

Our next series of BHNS simulations will involve spinning BHs.  Most
formation scenarios for BHNSs favor spinning BHs, especially systems
in which the BH spin and orbital angular momenta axes are nearly
aligned \cite{Schnittman}.  Since the ISCO for a prograde BH lies at a
smaller radius than that of a non-spinning BH, we expect that the
tidal disruption of the NS around these BHs occurs farther from the
ISCO.  Thus, spinning BHs would likely lead to a more massive disk,
but the magnitude of the effect and the scaling with respect to spin
will need to be determined via numerical calculations (as suggested in
\cite{RKLRasio}, for the $q \simeq 10$ cases).  Such
calculations will enable us to probe in depth which areas in phase
space are likely to serve as progenitors for SGRBs.

We find that the GW signal resulting from our BHNS coalescences is
attenuated at frequencies roughly equal to double the orbital
frequency at which tidal disruption begins, as one would expect,
confirming the fits described in \cite{TBFS07b}.  The
deviation between BHNS and BHBH inspiral is visible in the advanced
LIGO band for systems with mass ratios $q=3$ out to distances
$\gtrsim 100~{\rm Mpc}$, within which volume some population synthesis
calculations predict $\sim 1$ BHNS merger per year \cite{BBR}.  Should the
chirp mass determination, combined with higher order PN waveform phase
effects, allow for an independent determination of the component masses
of the binary, observation of the BHNS merger break frequency should
give a solid estimate of the NS radius.  Such effects are independent
of the discussion of disk formation, since the GW signal is strongly
suppressed after the onset of tidal disruption.

We have performed a series of calculations for the $q=3$ mass ratio
case (sequence~A), in which we vary only grid parameters to determine
the numerical resolution requirements for these BHNS mergers.  Gross
features of the hydrodynamics, such as the accretion rate onto the BH,
seem insensitive to the numerical resolution on the grid, at least for
this case, where the NS accretes fairly promptly.  The frequencies of
the waveforms at critical moments are similarly insensitive to
resolution.  Waveform amplitudes, on the other hand, vary by a larger
amount with respect to resolution, and are seen to be accurate only at
times when the constraint violations remain small.  At late times,
when they are largest, constraint violations are dominated by finite
boundary effects, which can be greatly reduced by enlarging the
physical extent of the grid.  Given the computational resource
requirements for these simulations, it may be extremely costly to
calculate waveforms accurate to a few percent using fisheye grids or
similar fixed-mesh refinement techniques.

We expect that high accuracy calculations spanning $\sim10$
orbits, as are currently performed in BHBH mergers, will require
us to use adaptive mesh refinement (AMR) techniques.  Our current
technique includes a single high resolution grid that
encompasses both the BH, NS, and surrounding strong-field region.
Outside of this region is a transition to a lower resolution grid
domain that extends to the outer grid boundary.
With AMR, we will be able to focus this high resolution entirely on the two
regions immediately surrounding the BH and NS.  With significantly
fewer gridpoints in the strong-field region of the grid, we will be
able to place more gridpoints in the low-resolution, weak-field region, 
thus extending our outer boundaries.  In many ways, a relativistic
hydrodynamics code with AMR will likely become a key tool for
simulating BHNS spacetimes.

\acknowledgments
We thank V.~Kalogera and S.~Hughes for useful conversations, and
the latter for providing a tabulated version of the advanced LIGO
noise sensitivity curve.  This paper was supported in part by NSF
Grants PHY02-05155, PHY03-45151, and PHY06-50377 as well as NASA
Grants NNG04GK54G and NNX07AG96G to University of Illinois at
Urbana-Champaign, and NSF Grant PHY04-56917 to Bowdoin College.  JAF
was also supported in part by NSF Grant AST04-01533.  All simulations
were performed on the NCSA {\tt abe} cluster.

\bibliography{bhnsgr}

\end{document}